\documentclass[lettersize,journal]{IEEEtran}
\IEEEoverridecommandlockouts
\usepackage{multirow}
\usepackage{longtable}
\usepackage{tabularx}
\usepackage[table,xcdraw]{xcolor}
\usepackage[bottom]{footmisc}
\usepackage{empheq}
\usepackage{float}
\usepackage{cite}
\usepackage{orcidlink}
\usepackage{bm}
\usepackage{amsmath,amssymb,amsfonts}
\usepackage{graphicx}
\usepackage{textcomp}
\usepackage{booktabs}
\usepackage{chemformula}
\def\BibTeX{{\rm B\kern-.05em{\sc i\kern-.025em b}\kern-.08em
T\kern-.1667em\lower.7ex\hbox{E}\kern-.125emX}}
\usepackage{mwe}
\usepackage[utf8]{inputenc}
\usepackage[english]{babel}

\usepackage{algpseudocode}
\usepackage[linesnumbered,ruled,vlined]{algorithm2e}
\usepackage[font=small,format=hang,parskip=0pt]{caption}
\usepackage[font=small,skip=0pt]{caption}
\usepackage{balance}
\usepackage{url}
\usepackage[normalem]{ulem}
\usepackage{caption}
\usepackage{subcaption}
\usepackage{xspace}
\PassOptionsToPackage{colorlinks=false, pdfborder={0 0 0}, breaklinks}{hyperref}
\newcommand{\etal}{\emph{et al.\xspace}}
\newcommand{\ie}{\emph{i.e.}, }

\newcommand{\DQL}{\texttt{DQ-Ladder}\xspace}
\usepackage[inline]{enumitem}
\setlist[enumerate,1]{label=\textit{\alph*)}}
\usepackage[binary-units=true]{siunitx}
\DeclareSIUnit[number-unit-product = ]\pixel{P}
\usepackage{tikz}

\arrayrulecolor{black}
\usepackage{ragged2e} 
\begin{document}
\title{DQ-Ladder: A Deep Reinforcement Learning-based Bitrate Ladder for Adaptive Video Streaming}
\author{Reza~Farahani~\orcidlink{0000-0002-2376-5802}, Zoha Azimi~\orcidlink{0009-0007-5405-0643},
Vignesh V Menon~\orcidlink{0000-0003-1454-6146},\\
Hermann Hellwagner~\orcidlink{0000-0003-1114-2584}, Radu Prodan~\orcidlink{0000-0002-8247-5426}, Schahram Dustdar~\orcidlink{0000-0001-6872-8821} and~Christian Timmerer~\orcidlink{0000-0002-0031-5243}
\thanks{
Reza Farahani, Zoha Azimi, Hermann Hellwagner, and Christian Timmerer are with the ITEC department, University of Klagenfurt, Austria (firstname.lastname@aau.at). 
Vignesh V Menon is with the University of Technology Cottbus-Seifert, Germany (e-mail: Vignesh.Menon@b-tu.de).
Radu Prodan is with the University of Innsbruck, Austria (radu.prodan@uibk.ac.at). 
Schahram Dustdar is with the Research Division of Distributed Systems at the TU Wien University of Vienna, Austria (dustdar@dsg.tuwien.ac.at). 
}}
\maketitle
\begin{abstract}
Adaptive streaming of segmented video over HTTP typically relies on a predefined set of bitrate-resolution pairs, known as a \emph{bitrate ladder}. However, fixed ladders often overlook variations in
content and decoding complexities, leading to suboptimal
trade-offs between encoding time, decoding efficiency, and video
quality. This article introduces \DQL, a deep reinforcement
learning (DRL)-based scheme for constructing time- and quality-aware bitrate ladders for adaptive video streaming applications.
\DQL employs predicted decoding time, quality scores, and bitrate levels per segment as inputs to a Deep Q-Network (DQN) agent, guided by a weighted reward function of
decoding time, video quality, and resolution smoothness. We leverage machine learning models to predict decoding time, bitrate level, and objective quality metrics (VMAF, XPSNR), eliminating the need for exhaustive encoding or quality metric computation. We evaluate \DQL using the \textit{Versatile Video Coding} (VVC) toolchain (VVenC/VVdeC) on \num{750} video sequences across six Apple HLS-compliant resolutions and \num{41} quantization parameters. Experimental results against four
baselines show that \DQL achieves BD-rate reductions of at least \qty{10.3}{\percent} for XPSNR compared to the HLS ladder, while reducing decoding time by \qty{22}{\percent}. \DQL shows significantly lower sensitivity to prediction errors than competing methods, remaining robust even with up to \qty{20}{\percent} noise.
\end{abstract}
%%%
%%%

\section{Introduction}\label{sec:Introduction}
Video traffic, mainly generated by video-on-demand and live streaming, dominates global Internet traffic, increasing demands for high-quality, low-latency, and sustainable content delivery~\cite{AppLogic25, bentaleb2025toward}. Among streaming technologies, \emph{HTTP Adaptive Streaming} (HAS), standardized by MPEG Dynamic Adaptive
Streaming over HTTP (DASH)~\cite{DASH_ref} and HTTP Live Streaming
(HLS)~\cite{HLS}, has become the dominant video delivery mechanism. HAS splits video content into segments, each encoded at multiple bitrates and resolutions, known as
representations forming a \emph{bitrate ladder}, stored on HTTP
servers. HAS clients employ \emph{adaptive bitrate} algorithms to select and retrieve each segment from the
ladder in an appropriate representation, optimizing playback
quality~\cite{bentaleb2022bob}. However, the growing demand for high-resolution content and advances in video codecs have expanded the number of representations in the ladder to accommodate diverse device capabilities~\cite{netflix_paper}.

Traditionally, bitrate ladder construction has relied on fixed rules that ignore content-specific characteristics affecting visual quality and encoding efficiency~\cite{lebreton2023quitting}. Netflix’s per-title encoding~\cite{netflix_paper} marked a shift toward content-aware encoding by analyzing each video’s rate–distortion behavior to derive improved ladders using perceptual quality metrics such as Video Multimethod Assessment Fusion (VMAF)~\cite{VMAF}. Building on this paradigm, several recent works~\cite{gnostic,res_pred_ref1,premkumar2024quality} determine resolutions within a target bitrate range to further improve compression efficiency and perceptual quality. However, these approaches typically overlook decoding time, which is increasingly critical for resource-constrained end-user devices. Consequently, balancing visual quality and decoding time remains a key challenge in ladder construction. Conventional heuristic-based approaches~\cite{azimi2024decoding,katsenou2024decoding} optimize resolution and quantization parameters (QPs) independently at each target bitrate, ignoring how early decisions influence later trade-offs such as quality consistency, decoding complexity growth, and smooth resolution transitions across the ladder~\cite{telili2023bitrate}. In contrast, \emph{deep reinforcement learning} (DRL) formulates ladder construction as a sequential decision problem, enabling a global policy that anticipates long-term quality and complexity trade-offs and coordinates decisions across the entire ladder.

This article proposes \DQL, a DRL-based framework that balances quality, decoding time, and resolution smoothness. It exploits video complexity features and incorporates lightweight machine learning (ML) predictors within a modular architecture to predict the expected bitrate, decoding time, and objective quality (measured via VMAF and XPSNR) for unseen video sequences. These predictions guide a \emph{Deep Q-Learning} (DQN)
agent to explore the multi-objective trade-off space. Extensive experiments on a dataset of \num{750} diverse real-world video sequences with varying content complexities demonstrate that \DQL outperforms four state-of-the-art methods, achieving 
BD-rate savings of at least \qty{10.9}{\percent} (PSNR) and 
\qty{10.3}{\percent} (XPSNR), while lowering decoding and encoding time by up to \qty{22}{\percent} and \qty{31}{\percent}. \DQL is robust to prediction errors, maintaining stable quality–decoding trade-offs even when rate–distortion and decoding-time estimates are perturbed by up to \qty{20}{\percent} noise. 

The contributions of \DQL are four-fold:
%%%
\begin{enumerate*}[label=(\emph{\roman*)}]
\item\emph{Modular architecture:} We design a DRL-based architecture leveraging DQN for bitrate ladder construction, jointly optimizing quality and decoding complexity.
\item\emph{Metric prediction:} We leverage lightweight ML models to predict bitrate, decoding time, and quality metrics, enabling online content-adaptive ladder generation.
\item\emph{Problem formulation:} We design a weighted reward function to efficiently explore the \textit{bitrate–quality–decoding time} trade-off space, incorporating constraints and penalties to guide decision-making.
\item\emph{Comprehensive evaluation:} We conduct extensive experiments using \num{750} real video sequences with various encoding configurations and the VVC codec, showing the superiority of \DQL over four baselines in terms of rate-distortion performance, decoding time reduction, and encoding speed-up.
\end{enumerate*}

The article has eight sections. Section~\ref{sec:RelatedWork} reviews related work. Section~\ref{sec:SM} formulates the bitrate ladder construction problem as a Markov decision process (MDP). Section~\ref{sec:arch} presents its architecture, followed by the execution workflow of \DQL in Section~\ref{sec:alg}. Section~\ref{sec:setup} describes the dataset, prediction models, and experimental setup. Section~\ref{sec:EResult} analyzes the evaluation results. Finally, Section~\ref{sec:conclusion} concludes the article and discusses future directions.

\section{Related Work}
\label{sec:RelatedWork}
While Mean Opinion Score (MOS) provides the most accurate perceptual quality assessment, its cost and subjectivity make it impractical for real-time applications, motivating learning-based approaches for MOS prediction~\cite{li2024perceptual,zou2017event, hands2004basic}. Several works employed decision trees (DT)~\cite{hameed2016decision}, multi-layer perceptron~\cite{begluk2018machine}, or federated learning~\cite{ickin2021qoe}, leveraging encoding parameters, network conditions (delay and packet loss), and user data for MOS prediction. Beyond MOS, PSNR~\cite{psnr_ref}, SSIM~\cite{ssim_ref}, and VMAF~\cite{VMAF} have been widely adopted for content-aware video streaming and encoding optimization. 
Amirpour~\etal~\cite{amirpour2025vqm4has} present VQM4HAS, a regression model that approximates VMAF for HAS representations using low-cost features and encoder logs, avoiding explicit VMAF computation at scale.
Although XPSNR seems more artifact-sensitive and computationally efficient than VMAF ~\cite{wien_xpsnr_vs_vmaf, itu_xpsnr_vs_vmaf, xpsnr_vs_vmaf}, predictive modeling remains unexplored.

Kranzler~\etal~\cite{kranzler2020decoding} designed a bitstream-based decoding energy prediction model using multivariate linear regression, integrated into their proposed encoder for energy-aware rate-distortion optimization. Boujida~\etal~\cite{boujida2023decoding} proposed an ML-based VVC per-frame decoding time prediction using extra trees regression (ETR) with features extracted from the bitstream, \ie coding unit size, split depth, motion vectors, enabling adaptive decoding control. Ghasempour~\etal~\cite{ghasempour2025real} modeled decoding energy using video resolution and framerate as features, leveraging a lightweight lookup table approach for fast estimation in live streaming. Boujida~\etal~\cite{cabarat2024pictures} extended their per-frame VVC decoding time model by integrating lightweight temporal feedback (past decoding times) and feature selection, enabling accurate multi-resolution prediction for dynamic voltage and frequency scaling (DVFS)-based live streaming adaptation.
Existing works rely on codec-dependent syntax or full bitstream parsing, limiting low-latency online use; their generalization to unseen content and diverse encoding settings remains largely unexplored.

Katsenou~\etal~\cite{gnostic} provided a content-driven bitrate ladder for HEVC-encoded video sequences using regression-based prediction models. Bhat~\etal~\cite{res_pred_ref1} employed RF models to predict the best resolution for each scene, while Mueller~\etal~\cite{mueller2022context} combined convolutional neural networks, XGBoost, and MLP for content complexity analysis. Menon~\etal~\cite{menon2023jnd} used random forest (RF) models to predict VMAF and select bitrate-resolution in a Just Noticeable Difference~\cite{jnd_ref} (JND)-aware encoding scheme. Azimi~\etal~\cite{azimi2024decoding} integrated XGBoost models to estimate encoding time and quality metrics, enabling decoding-aware bitrate ladder construction. Lebreton~\etal~\cite{lebreton2023quitting, lebreton2020predicting} proposed ladder generation based on user quitting probability to align quality with user engagement. 
Rajendran~\etal~\cite{rajendran2024} used Pareto-front analysis to predict optimized framerates for constructing energy-efficient ladders. 
Zhao~\etal~\cite{zhao2024efficient} proposed a per-shot bitrate ladder construction method that encodes a small subset of resolution-bitrate-preset points and predicts the remaining rate-distortion curves using curve fitting and ML models (e.g., RF, support vector machines, neural networks). 
Existing methods focus on quality alone, neglecting decoding complexity and resolution smoothness, and requiring exhaustive pre-encoding.

\section{\DQL Model}\label{sec:SM}
We formulate bitrate ladder construction as a \textit{Markov decision process} (MDP). 
%%%%%%%%%%
\subsubsection{Environment}
We define the environment as the video encoding system operated by the video providers, where each video is split into segments \mbox{$\mathcal{S}=\{s_1, s_2,\dots, s_m\}$}, each encoded at a set of target bitrates \mbox{$\mathcal{TB}=\{tb_1,tb_2,…,tb_n\}$}. For each target bitrate \mbox{$tb_i\in\mathcal{TB}$}, the DRL agent selects a resolution \mbox{$r_i\in\mathcal{R}$} and quantization parameter \mbox{$qp_i\in\mathcal{QP}$} from predefined sets. The environment provides three essential feedback signals per segment based on lightweight predictive models: decoding time \mbox{$\widehat{\mathcal{T}}=f_1(E_Y, h, L_Y, \mathcal{R},\mathcal{QP})$}, quality score \mbox{$\widehat{\mathcal{Q}} = f_2(E_Y, h, L_Y, \mathcal{R}, \mathcal{QP})$}, and bitrate level \mbox{$\widehat{\mathcal{B}} = f_3(E_Y, h, L_Y, \mathcal{R}, \mathcal{QP})$}, where \mbox{$f_1$, $f_2$, and $f_3$} are the corresponding predictive functions. Here, $E_Y$ shows luma texture complexity, $h$ denotes temporal complexity, and $L_Y$ represents luma brightness extracted from the video segment. 
%%%%%%%%%%%%
\subsubsection{State space}
We define the state space \mbox{$\mathbb{S}=\{st_1,st_2,\dots,st_k\}$} 
with total size \mbox{$|\mathcal{S}|\cdot|\mathcal{TB}|$}, encompassing all possible combinations of video segments and target bitrate levels. We define the state \mbox{$st_i\in\mathbb{S}$} as $st_i = \left[tb_i, \widehat{b}_{i-1}, \widehat{q}_{i-1}, \widehat{t}_{i-1} \right]$
, where it captures the current target bitrate $tb_i$ for each segment and the predicted outcomes of the previous encoding step $i-1$, i.e., bitrate \mbox{$\widehat{b}_{i-1}\in\widehat{\mathcal{B}}$}, quality \mbox{$\widehat{q}_{i-1}\in\widehat{\mathcal{Q}}$}, and decoding time \mbox{$\widehat{t}_{i-1}\in\widehat{\mathcal{T}}$}, allowing the agent to reason encoding behavior without incurring the high computational cost of real encoding and metric computation.
%%%%%%%%%%%%
%%%%%%%%%%%%%%%%%%%%%
\begin{figure*}[!t]
	\centering
	\includegraphics[width=.85\linewidth]{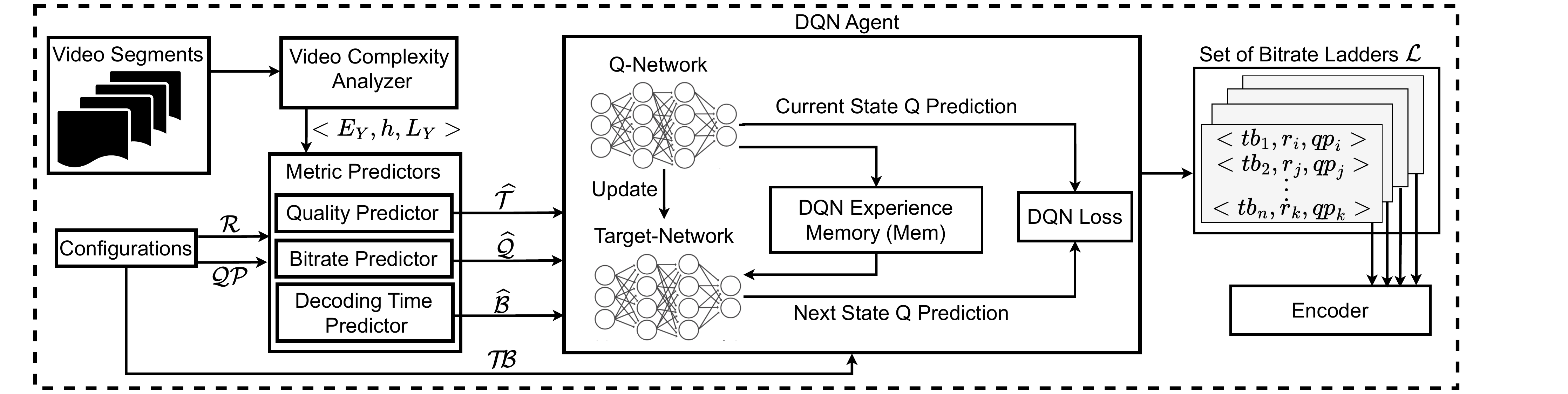}
	\caption{\DQL system architecture.}
	\label{arch}
\end{figure*}
%%%%%%%%%%%%%%%%%%%%%
\subsubsection{Action space}
We define the action set \mbox{$\mathcal{A}=\{a_1,a_2, \dots, a_n\}$}, where each \mbox{$a_i\in\mathcal{A}$} specifies an encoding configuration for target bitrate $tb_i\in\mathcal{TB}$ defined as $a_i = \left[r_i, qp_i\right]$,
where \mbox{$r_i\in\mathcal{R}$} is the chosen resolution and \mbox{$qp_i\in\mathcal{QP}$} is the selected quantization parameter. The overall action space \mbox{$\mathbb{A}=\{\mathcal{A}_1,\dots,\mathcal{A}_k\}$} is the union of all possible actions across video segments \mbox{$\mathcal{S}$} and target bitrates  \mbox{$\mathcal{TB}$}, representing the full set of feasible encoding decisions during ladder construction.
\subsubsection{Reward function}
The reward function guides the agent in building an efficient bitrate ladder by maximizing quality, minimizing decoding time, and preserving visual stability across segment encodings. After selecting action \mbox{$a_i=[r_i,qp_i]$} for segment \mbox{$s_j\in\mathcal{S}$} at target bitrate \mbox{$tb_i\in\mathcal{TB}$}, the agent receives a reward \mbox{$Rew_i$} formulated as:
\begin{equation} 
Rew_i =\lambda_1\cdot \widehat{\bar{q}}_i - \lambda_2 \cdot \widehat{\bar{t}}_i - \lambda_3 \cdot \delta_i.
\label{rew}
\end{equation}
where \mbox{$\widehat{\bar{q}}_i$} and \mbox{$\widehat{\bar{t}}_i$} denote the normalized predicted quality and decoding time, calculated using min-max normalization: 
\begin{equation} 
\widehat{\bar{q}}_i = \frac{\widehat{q}_i - \widehat{q}_{\min}}{\widehat{q}_{\max} - \widehat{q}_{\min}}, \quad \widehat{\bar{t}}_i = \frac{\widehat{t}_i - \widehat{t}_{\min}}{\widehat{t}_{\max} - \widehat{t}_{\min}}.
\end{equation} 
where \mbox{$\widehat{q}_{\min}$, $\widehat{q}_{\max}$, $\widehat{t}_{\min}$}, and \mbox{$\widehat{t}_{\max}$} represent the minimum and maximum predicted quality and decoding time values across the candidate actions.
To penalize frequent resolution switches between segments and avoid unnecessary visual fluctuation, we define the resolution change penalty $\delta_i$ as:
\begin{equation} 
\label{eq:penalty}
\delta_i = \begin{cases} 1, & \text{if } r_i \neq r_{i-1}; \\ 
0, & \text{otherwise}.
\end{cases}
\end{equation}
where \mbox{$r_{i-1}\in\mathcal{R}$} is the resolution selected in the previous encoding step $i-1$. The penalty term ensures visual consistency across segments, avoiding unnecessary resolution fluctuations. The $\lambda$ coefficients control the trade-off among quality maximization, decoding efficiency, and resolution stability and are empirically selected based on system preferences.
%%%%%%%%%%%%
\subsubsection{Encoding constraints}
We impose two essential constraints to guarantee bitrate and quality control in the constructed ladder.
First, the predicted bitrate $\widehat{b}_i$ of the selected action \mbox{$a_i=[r_i, qp_i]$} must not exceed the target bitrate threshold $tb_{max}$, ensuring bitrate-constrained encoding per segment:
\begin{equation} 
\widehat{b}_i \leq tb_{\max}.
\end{equation}
Second, we enforce quality monotonicity across increasing target bitrates to avoid quality degradation at higher rates:
\begin{equation} \label{eq:monoton}
tb_i > tb_{i-1} \Rightarrow \widehat{q}_i \geq \widehat{q}_{i-1}.
\end{equation}
which guarantees an ordered bitrate ladder where higher bitrate levels always provide equal or better objective quality.

\section{\DQL Architecture}\label{sec:arch}
We design a modular architecture for \DQL to enable efficient video segment-level bitrate ladder construction, as illustrated in Fig.~\ref{arch}. The architecture integrates three main modules: \textit{(1)} video complexity analyzer, \textit{(2)} lightweight ML-based metric predictors, and \textit{(3)} a DQN agent.
Given a set of video segments $\mathcal{S}=\{s_1, s_2, \dots, s_m\}$, the video complexity analysis extracts spatial, temporal, and brightness features from each segment, and considers the luma-based features \mbox{$\langle E_Y, h, L_Y \rangle$}. These features, combined with candidate encoding parameters \mbox{$(\mathcal{R}, \mathcal{QP})$}, are provided to three independent predictors \mbox{$f_1, f_2$ and $f_3$}, estimating decoding time set $\widehat{\mathcal{T}}$, quality set $\widehat{\mathcal{Q}}$, and bitrate set $\widehat{\mathcal{B}}$ without real encoding overhead. The quality predictor is metric-agnostic. Although we train separate predictors for XPSNR and VMAF (Section~\ref{ml-impl}), the model used at inference time is selected according to the service provider’s optimization objective. For example, when optimizing VMAF, it instantiates the VMAF predictor, and the DQN agent maximizes rewards concerning that metric. The DQN agent interacts with the environment to construct the segment-specific ladder using three main steps:
%%%
\subsubsection{DQN agent operation}  
The agent models the state-action value function \mbox{$Q(st_i, a_i)$} via a neural network and iteratively refines it using observed experience. At each decision step, for a given segment \mbox{$s_j\in\mathcal{S}$} and target bitrate \mbox{$tb_i\in\mathcal{TB}$}, the agent observes the current state \mbox{$st_i=[tb_i,\widehat{b}_{i-1},\widehat{q}_{i-1},\widehat{t}_{i-1}]$}, capturing the current target bitrate $tb_i$ and the outcomes of the previous encoding step $(i-1)$. This design enables fast decision-making by leveraging learned predictions, avoiding real encoding computation while retaining awareness of past encoding behavior for stability. Based on this state, the agent selects an action \mbox{$a_i=[r_i, qp_i]$} from the action space $\mathcal{A}$, specifying the resolution \mbox{$r_i\in\mathcal{R}$} and quantization parameter \mbox{$qp_i\in\mathcal{QP}$} to encode the segment $s_j$ at $tb_i$.
The environment then returns the predicted metrics $(\widehat{b}_i,\widehat{q}_i,\widehat{t}_i)$ for the chosen configuration, used to compute the reward $Rew_i$ based on the multi-objective reward function defined in Eq.~\ref{rew}.
To learn a policy that optimizes long-term cumulative reward, the agent minimizes the \emph{DQN loss}, defined as the mean squared error between the Q-value predicted by the Q-network for the current state–action pair and a target value computed using the target network. The Q-network parameters are updated according to the Bellman optimality equation (Eq.~(\ref{Bellman})):
\begin{align}
Q(st_i,a_i) \leftarrow Rew_i + \gamma\cdot \max_{a'} Q\left(st_{i+1},a'\right).
\label{Bellman}
\end{align}
where $a'$ denotes all possible actions at the next state $st_{i+1}$, and $\gamma$ is the discount factor that controls future reward impact.
\subsubsection{DQN experience memory} The agent maintains an experience replay memory $Mem$ to enhance learning stability and avoid overfitting to recent decisions. After selecting action \mbox{$a_i=\left[r_i, qp_i\right]$} in state $st_i$ and observing reward $Rew_i$ and next state $st_{i+1}$, the agent stores the tuple \mbox{$\left(st_i, a_i, Rew_i, st_{i+1}\right)$} in $Mem$. To update the Q-network, the agent uniformly samples mini-batches from $Mem$, enabling diverse learning from past encoding experiences across different segments in $\mathcal{S}$. 
\subsubsection{DQN output} The decision process continues iteratively for all \mbox{$tb_i\in\mathcal{TB}$}, constructing the segment-level ladder \mbox{$\mathcal{L}_j=\left\{\left(tb_1, r_i, qp_i\right), \dots, \left(tb_n, r_k, qp_k\right)\right\}$} for $s_j$. The final output is the complete set of bitrate ladders $\mathcal{L}=\{\mathcal{L}_1, \mathcal{L}_2, \dots, \mathcal{L}_m\}$ covering all segments in $\mathcal{S}$, subsequently passed to the encoder for final processing.

\section{\DQL Execution Workflow}
\label{sec:alg} 
The operational workflow of \DQL includes two phases: \textit{(1) DQN training} conducted provider-side using realistic encoding feedback and \textit{(2) DQN inference} for ladder construction using lightweight predictions. 
%Table~\ref{Tab:notation} summarizes the main notations used for the algorithms.
%%%%%%%%%%
\subsubsection{DQN training}
Alg.~\ref{alg:train} shows the training procedure of the DQN agent, performed on a provider-side video dataset to learn an effective bitrate ladder construction policy. The agent interacts with a real encoding environment, where each segment \mbox{$s_j\in\mathcal{S}$} is encoded with different configurations \mbox{$(r_i, qp_i)$} to collect actual decoding time $t_i$, quality $q_i$, and bitrate $b_i$ values. At each training epoch $e\in\mathcal{EP}$ (line 4) and decision step for segment $s_j$ and target bitrate \mbox{$tb_i \in \mathcal{TB}$} (lines 5 -6), the agent observes the state \mbox{$st_i=[tb_i, b_{i-1}, q_{i-1}, t_{i-1}]$} using prior encoding results (line 7). The agent selects an action \mbox{$a_i = [r_i, qp_i]$}, specifying the resolution $r_i$ and quantization parameter $qp_i$, using an $\epsilon$-greedy strategy (lines 8–11). With probability $\epsilon$, a random action is chosen to explore new configurations; otherwise, with probability $1-\epsilon$, the action with the highest current Q-value is selected for exploitation.

The selected action triggers real encoding (line 12), from which the measured bitrate $b_i$, quality $q_i$, and decoding time $t_i$ are obtained and used to compute the reward $Rew_i$ (lines 13–14).
The next state $st_{i+1}$ is then observed (line 15), and the experience tuple $(st_i, a_i, Rew_i, st_{i+1})$ is then stored in the experience replay memory $Mem$ (line 16). Next, the Q-network parameters $\theta$ are updated using mini-batches from $Mem$ (line 17), minimizing the temporal difference loss between the predicted and target Q-values (lines 18-19). To stabilize learning, the target network parameters $\theta^{-}$ are periodically synchronized with $\theta$ every $C$ steps (line 20). 
Next, the exploration rate $\epsilon$ is linearly decayed toward $\epsilon_{\text{end}}$ (line 21) and clamped to avoid falling below $\epsilon_{\text{end}}$ (line 22). Finally, the training proceeds over subsequent epochs (line 23) until convergence. Finally, the output is the trained Q-function with parameters $\theta$ used for inference (line 24). 

We note that the DQN training procedure is codec-agnostic and independent of specific reward weight values. Codec-specific effects are implicitly captured by the measured bitrate, quality, and decoding time values obtained during training, while the reward weights reflect the provider's optimization preferences. Thus, adapting \DQL to a different codec or service objective does not require changes to the learning formulation or network architecture, but only retraining with the corresponding measurements and updated reward weights.
%%%%%%%%%%%%%%%%%%%%%%%%%%%%%%
\begin{algorithm}[!t]
{\fontsize{7pt}{7pt}\selectfont
\caption{DQN Training Algorithm.}
\label{alg:train}
\KwIn{Video segments $\mathcal{S}$, Target bitrates $\mathcal{TB}$, Resolutions $\mathcal{R}$, QPs $\mathcal{QP}$, discount factor $\gamma$, exploration rate start $\epsilon_{\text{start}}$, exploration rate end $\epsilon_{\text{end}}$, batch size $BA$, memory $Mem$, Epoch size $\mathcal{EP}$}
\KwOut{Trained Q-network parameters $\theta$}
Initialize replay memory $Mem$ \\
Initialize network $Q(st,a;\theta)$ and target network $\tilde{Q}(st,a;\theta^{-}) \leftarrow \theta$\\
Initialize $\epsilon \leftarrow \epsilon_{\text{start}}$\\
\For{$e \leq \mathcal{EP}$}{
    \For{$s_j \in \mathcal{S}$}{
        \For{$tb_i \in \mathcal{TB}$}{
            $st_i\leftarrow[tb_i, b_{i-1}, q_{i-1}, t_{i-1}]$\\
            \eIf{Rand() $< \epsilon$}{
                $a_i\leftarrow$ Rand$([r_i, qp_i])$\\
            }{
                $a_i \leftarrow \arg\max_{a} Q(st_i,a;\theta)$\\
            }
            Execute($a_i$)\\
            Measure($b_i, q_i, t_i$)\\
            $Rew_i\leftarrow\lambda_1 \cdot \bar{q}_i - \lambda_2 \cdot \bar{t}_i - \lambda_3 \cdot \delta_i$\\
            Observe($st_{i+1}$)\\
            $Mem\leftarrow(st_i, a_i, Rew_i, st_{i+1})$\\
            Sample($Mem, BA$)\\
            $y_i\leftarrow Rew_i + \gamma \max_{a'} \tilde{Q}(st_{i+1},a';\theta^{-})$\\
            Update($\theta$, $(y_i - Q(st_i,a_i;\theta)))^2$\\
            Every $C$ steps: $\theta^{-} \leftarrow \theta$\\
        }
    }
    $\epsilon \leftarrow \epsilon_{\text{start}} - (\epsilon_{\text{start}} - \epsilon_{\text{end}}) \cdot \frac{e}{\mathcal{EP}}$\\
    $\epsilon \leftarrow \max(\epsilon_{\text{end}}, \epsilon)$\\
    $e \leftarrow e + 1$\\
}
Return ($Q(st, a; \theta))$}
\end{algorithm}
%%%%%%%%%%%%%%
%%%%%%%%%
\begin{algorithm}[!t]
{\fontsize{7pt}{7pt}\selectfont
\caption{DQN Inference Algorithm.}
\label{alg-inf}
\KwIn{Video segments $\mathcal{S}$, Target bitrates $\mathcal{TB}$, Resolutions $\mathcal{R}$, QPs $\mathcal{QP}$, trained $Q(st,a;\theta)$, predictors $f_1, f_2, f_3$}
\KwOut{Bitrate ladder $\mathcal{L} = \{\mathcal{L}_1, \dots, \mathcal{L}_m\}$}
\For{$s_j \in \mathcal{S}$}{
    Extract ($E_Y, h, L_Y$)\\
    
    \For{$tb_i\in\mathcal{TB}$}{
        Observe $st_i\leftarrow[tb_i, \widehat{b}_{i-1}, \widehat{q}_{i-1}, \widehat{t}_{i-1}]$\\
        Select $a_i\leftarrow \arg\max_{a} Q(st_i, a; \theta)$\\
        ($\widehat{t}_i, \widehat{q}_i, \widehat{b}_i$) $\leftarrow$ ($f_1, f_2, f_3$)\\
        $\mathcal{L}_j\leftarrow(tb_i, r_i, qp_i)$ \\
    }
    Return ($\mathcal{L}_j$)
}}
\end{algorithm}
%%%%%%%
\subsubsection{DQN inference}
Alg.~\ref{alg-inf} shows the DQN online inference procedure.
The trained DQN agent operates without real encoding, leveraging lightweight content features and ML-based prediction models. For each segment \mbox{$s_j\in\mathcal{S}$} (line 1), the video complexity analyzer extracts content features $\langle E_Y, h, L_Y \rangle$ (line 2). At each target bitrate \mbox{$tb_i\in\mathcal{TB}$} (line 3), the agent observes the current state $st_i=[tb_i, \widehat{b}_{i-1}, \widehat{q}_{i-1}, \widehat{t}_{i-1}]$ (line 4) using the target bitrate and predicted metrics from the previous step. The action $a_i=[r_i, qp_i]$ is then selected by maximizing the learned Q-function (line 5). The environment returns the predicted outcomes $(\widehat{b}_i, \widehat{q}_i, \widehat{t}_i)$ using the predictors $f_1, f_2,$ and $f_3$ (line 6), avoiding costly real encoding. This process iterates over all target bitrates to construct the segment-specific bitrate ladder $\mathcal{L}_j$ (line 8). The corresponding ladder $\mathcal{L}_j$ is finally returned (line 8), making it available for encoding.
%%%%%%%
\subsubsection{Computational complexity analysis}
The offline DQN training (Alg.~\ref{alg:train}) has a time complexity of $\mathcal{O}(\mathcal{EP} \cdot |\mathcal{S}| \cdot |\mathcal{TB}| \cdot (C_{enc} + BA \cdot C_{NN}))$, where $\mathcal{EP}$ is the number of epochs, $|\mathcal{S}|$ is the number of video segments, $|\mathcal{TB}|$ is the number of target bitrates, $C_{enc}$ is the cost of real encoding per action (executed offline), $BA$ is the mini-batch size, and $C_{NN}$ is the neural network computation cost for forward and backward passes. The online inference (Alg.~\ref{alg-inf}) operates with complexity of $\mathcal{O}(|\mathcal{S}| \cdot |\mathcal{TB}| \cdot (C_{NN} + C_{pred}))$, where $C_{pred}$ is the worst time of the lightweight prediction models $f_1, f_2, f_3$. 

\section{Evaluation Setup}
\label{sec:setup}
This section details the evaluation setup of \DQL. We conducted all experiments on a server with a \num{128}-core Intel Xeon Gold CPU and two NVIDIA Quadro GV100 GPUs. 
%%%
\subsection{Dataset explanation and analysis}
\begin{figure}[!t]
    \centering
\includegraphics[width=.7\linewidth]{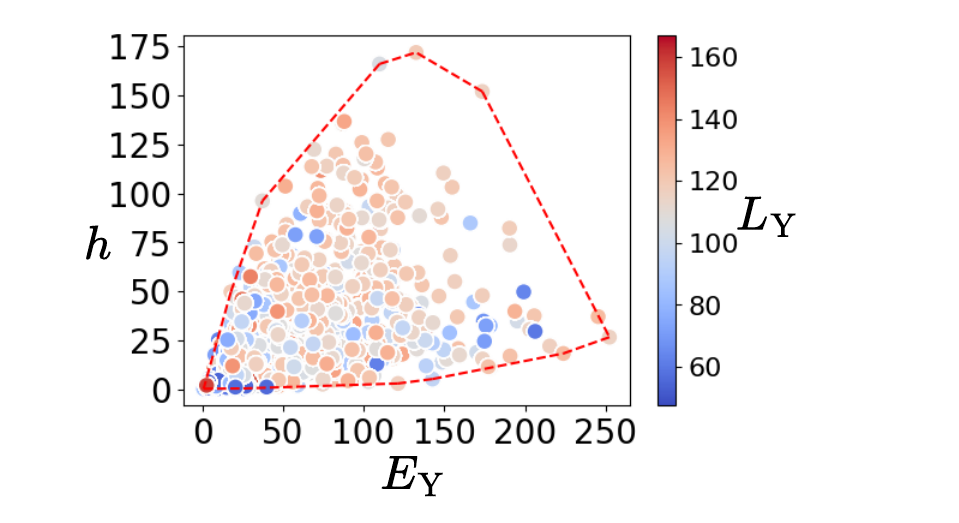}
    \caption{Content complexity feature distribution in the dataset.}
    \label{fig:dataset1}
\end{figure}

We use \num{750} ultra-high-definition (UHD) video sequences from the Inter-4K dataset~\cite{inter4k_ref}, spanning diverse content types such as nature, urban, sports, and synthetic scenes. Each sequence was encoded at \num{60}fps using \texttt{VVenC v1.11}~\cite{vvenc_ref} with the \texttt{faster} preset. As summarized in Table~\ref{tab:video_param}, the encoding setup covers six resolutions \mbox{$\mathcal{R} = \{360, 540, 720, 1080, 1440, 2160\}$p} and \num{41} quantization parameters \mbox{$\mathcal{QP}=\{10, \dots,50\}$}. For every encoded version, we recorded the actual bitrate $b_i$, encoding time, and the decoding time $t_i$ using \texttt{VVdeC v2.3.0}~\cite{VVdeC_ref}. Objective quality scores $q_i$ were measured using XPSNR and VMAF, providing ground-truth labels for training the DQN agent. Spatiotemporal complexity features were extracted using \texttt{VCA v2.0}~\cite{vca_ref}.
%%
% \begin{figure}[t]
%     \centering
%     \begin{subfigure}{0.49\columnwidth}
%     \centering    
%     \includegraphics[width=\linewidth]{Figures/Dataset2.pdf}
%     \caption{}
%     \label{fig:dataset2}    
%     \end{subfigure}
%     \begin{subfigure}{0.49\columnwidth}
%     \centering
%     \includegraphics[width=\linewidth]{Figures/Dataset3.pdf}
%     \caption{}
%     \label{fig:dataset3}
%     \end{subfigure}
%     \caption{Impact of (a) resolution variations, and (b) QP variations on encoding/decoding time, bitrate levels, and quality scores.}
% \end{figure}
%%
%%%
\begin{table}[!t]
\centering
\caption{Encoding and decoding configuration parameters.}
\label{tab:video_param}
\fontsize{7pt}{7pt}\selectfont
\begin{tabular}{cc}
\toprule
\textit{Parameter} & \textit{Configuration}   \\ \hline
$\mathcal{R}$                  
& $\{360, 540, 720, 1080, 1440, 2160\}$p \\ \hline
$\mathcal{QP}$                  
& $\{10, 11, 12,\ldots, 50\}$   \\ \hline
$\mathcal{TB}$  & \begin{tabular}[c]{@{}c@{}}$\{145, 300, 600, 900, 1600, 2400, 3400,$ \\ $4500, 5800, 8100, 11600, 16800\}$~Kbps\end{tabular} \\ \hline
Encoder            & VVenC, 4 CPU threads \\ \hline
Decoder            & VVdeC, 4 CPU threads  \\  
\bottomrule
\end{tabular}
\end{table}
%%%
We analyzed the dataset characteristics to highlight the content diversity.
Fig.~\ref{fig:dataset1} shows the distribution of video sequences in the $(E_Y, h)$ feature space, highlighting the diversity of spatial and temporal complexity, with color indicating variations in luminance $L_Y$. 
%%%%%%%%%%%%%%%%%%%%%%%%%%%%%%
\begin{table*}[ht!]
    \centering
    \begin{minipage}[b]{0.4\linewidth}
        \centering
        \caption{Hyperparameter search space explored for ML-based prediction models.}
        \label{tab:Predictors_hyperp}
        \fontsize{7pt}{7pt}\selectfont
        \begin{tabular}{ll}
        \toprule 
        \textit{Predictive model} & \textit{Explored hyperparameters} \\ 
        \midrule
        LR & None  \\ \hline
        RF, ETR& $n_{trees}\in\{50,100,200,300\}$,\\& $d_{max}\in\{3,5,7,10\}$\\ \hline
        AdaBoost, LightGBM & $n_{trees}\in\{50, 100, 200, 300\}$ \\ 
        &  $\eta\in\{0.01, 0.1, 0.5\}$\\ \hline
        XGBoost & $n_{trees}\in\{50, 100, 200, 300\}$\\ 
        & $d_{max}\in\{3, 5, 7, 10\}$ \\ 
        & $\eta\in\{0.01, 0.1, 0.5\}$\\ \hline
        MLP & $h_{units}\in\{64, 128, 256, 512\}$   \\
        &$\eta\in[10^{-5}, 10^{-2}]$\\& 
        $bs\in\{32,64,128,256\}$\\
        \bottomrule
        \end{tabular}
    \end{minipage}\hfill
    \begin{minipage}[b]{0.3\linewidth}
        \centering
        \caption{DQN agent hyperparameters.}
        \label{tab:dqn_hyperparams}
        \fontsize{7pt}{7pt}\selectfont
        \begin{tabular}{ll}
            \toprule
            \textit{Hyperparameter} & \textit{Value} \\
            \midrule
            Learning rate ($\eta$) & 0.05, 0.005, 0.0005 \\
            Discount factor ($\gamma$) & 0.7 \\
            Batch size ($BA$)& 128 \\
            Replay memory size ($Mem$) & 10,000 \\
            Target network update interval ($C$) & 10 \\
            Initial exploration rate ($\epsilon_{start}$) & 0.8 \\
            Final exploration rate ($\epsilon_{end}$) & 0.05 \\
            Exploration decay schedule & Linear decay  \\
            Training episodes ($\mathcal{EP}$) & 2,000 \\
            \bottomrule
        \end{tabular}
    \end{minipage}\hfill
    \begin{minipage}[b]{0.2\linewidth}
        \centering
        \caption{HLS ladder.}
        \label{tab:hls}
        \fontsize{7pt}{7pt}\selectfont
        \begin{tabular}{cc}
        \toprule 
        \textit{Bitrate (kbps)} & \textit{Resolution} \\ \hline 
        \num{145}   & \num{360} \\    \hline
        \num{300}   & \num{360} \\    \hline
        \num{600}   & \num{540} \\     \hline
        \num{900}   & \num{540} \\     \hline
        \num{1600}   & \num{540} \\     \hline
        \num{2400}   & \num{720} \\     \hline
        \num{3400}   & \num{720} \\     \hline
        \num{4500}   & \num{1080} \\     \hline
        \num{5800}   & \num{1080} \\     \hline
        \num{8100}   & \num{1440} \\     \hline
        \num{11600}   & \num{2160} \\     \hline
        \num{16800}   & \num{2160} \\     
        \bottomrule         
        \end{tabular}    
    \end{minipage}
\end{table*}
%%%%%%%%%%%%%%%%%%%%%%%%%%%%%%
%%%
\subsection{ML-based metric prediction models} \label{ml-impl}
Given the high overhead of real encoding, decoding, and quality measurements, \DQL employs lightweight ML-based predictors for decoding time, quality, and bitrate, enabling fast ladder construction during online inference (Section~\ref{sec:alg}).
We evaluated seven well-known ML models covering four different categories:
\begin{enumerate*}
\item \emph{Linear Regression} (LR)~\cite{lr_ref} as a simple linear model;
\item \emph{Random Forest} (RF)~\cite{rf_ref} and \emph{Extra Trees Regression} (ETR)~\cite{geurts2006extremely} as tree-based ensemble models;
\item \emph{AdaBoost}~\cite{solomatine2004adaboost}, \emph{LightGBM}~\cite{ke2017lightgbm}, and \emph{XGBoost}~\cite{xgboost_ref} as gradient boosting-based ensembles;
\item \emph{Multi-Layer Perceptron} (MLP)~\cite{popescu2009multilayer} as a neural network with fully connected layers.
\end{enumerate*}

We partitioned the dataset into disjoint training (\qty{70}{\percent}) and testing (\qty{30}{\percent}) sets at the video level, ensuring that all segments from a given video belong exclusively to one set to avoid data leakage. Video identifiers were randomly shuffled with a fixed seed (seed = 42) to guarantee reproducibility and consistent splitting across the three prediction tasks. We applied GridSearchCV~\cite{grid} from Scikit-learn~\cite{scikit-learn} with five-fold cross-validation on the training set to search for hyperparameters in ensemble-based models. We performed an exhaustive grid search over predefined parameter spaces and retrained the best estimator on the training data. We used \qty{10}{\percent} of the training set as a validation set and applied \emph{Optuna}~\cite{optuna_2019} for hyperparameter optimization in the MLP model. 

We applied model-specific hyperparameter tuning based on the design flexibility of each predictor. While the LR model has no tunable parameters, ensemble-based models require optimization of the number of estimators ($n_{\text{trees}}$), maximum tree depth ($d_{\text{max}}$), and learning rate ($\eta$). We fine-tuned the number of hidden units ($h_{\text{units}}$), learning rate ($\eta$), and batch size ($bs$) for the MLP model. Table~\ref{tab:Predictors_hyperp} summarizes the explored hyperparameter search space per model. In our final evaluation, we used the best-performing model per target metric (decoding time, bitrate, XPSNR, VMAF), and reported the corresponding hyperparameter values in Section~\ref{prediction-results}.
%%%%%%%%
\subsection{DQN agent configurations and hyperparameters}
To approximate the Q-value function, we employ a multi-layer perceptron (MLP) that takes the four-dimensional state vector $s_t^i$ as input. All state features are normalized to the range $[0,1]$ using min–max scaling to improve stability and convergence. The network consists of three connected hidden layers with \num{256}, \num{128}, and \num{64} neurons, using ReLU activation functions~\cite{nair2010rectified}. The output layer is a linear fully connected layer of size $|\mathcal{A}|$, where each neuron represents the Q-value of a discrete action pair $(r_i, qp_i)$. The model is implemented in PyTorch with Kaiming (He) initialization~\cite{he2015delving} and optimized using Adam~\cite{kingma2014adam}.
The DQN agent is trained for \mbox{$\mathcal{EP}=2000$} episodes using experience replay with a memory size of \mbox{$|Mem|=10000$} and a mini-batch size of $BA=128$. The target network is synchronized every $C=10$ steps. We adopt an $\epsilon$-greedy exploration strategy, with $\epsilon$ linearly decayed from $0.8$ to \mbox{$\epsilon_{\min}=0.05$} over \num{2000} steps. Hyperparameters are tuned via a combination of Optuna-based optimization~\cite{optuna_2019} and manual refinement. The learning rate \mbox{$\eta\in\{0.05, 0.005, 0.0005\}$} (depending on the experiment) and discount factor $\gamma=0.7$. Table~\ref{tab:dqn_hyperparams} summarizes the final DQN agent hyperparameters.
%%%%%%%%
\subsection{Evaluation metrics}\label{metric-predic}
We use different evaluation metrics to assess \DQL performance and compare it with baselines.
\subsubsection{Prediction model metrics} We use the following standard regression metrics to evaluate the accuracy and generalization performance of the prediction models: 
\begin{enumerate*}
    \item\emph{Coefficient of determination} ($R^2$);
    \item\emph{Mean absolute error percentage} (MAE\%);
    \item\emph{Root mean squared error} (RMSE);
    \item\emph{Standard deviation of absolute errors} (SDAE);
    \item\emph{Inference time}.
\end{enumerate*}
\subsubsection{DQN agent convergence analysis} \label{metric-agent}
We evaluate the agent's learning behavior using the \emph{cumulative reward evolution}, which aggregates per-episode rewards over training steps to reflect overall policy performance; higher values indicate better learning progress.
\subsubsection{Bitrate ladder efficiency metrics}\label{metric-ladder}
We use the following metrics to evaluate the quality-efficiency trade-offs of the constructed bitrate ladders:
\begin{enumerate*}
\item\emph{Bjøntegaard delta rate} (BD-rate)~\cite{barman2024bj} measures the percentage bitrate increase required to achieve equivalent quality (XPSNR or VMAF); lower negative BD-rate indicates higher encoding efficiency.
\item\emph{Bjøntegaard delta quality} (BD-PSNR, BD-XPSNR, BD-VMAF) indicates the average quality difference over a bitrate range. Higher positive values indicate better coding efficiency at equivalent bitrates.
\item\emph{Bjøntegaard delta encoding time} (BD-EnTime) quantifies the average change in encoding time to achieve the same quality or bitrate as the baseline; lower values indicate faster encoding.
\item\emph{Bjøntegaard delta decoding time} (BD-DeTime) measures the average change in decoding time to achieve the same quality or bitrate as the baseline; lower values indicate faster decoding.
\item\emph{Resolution switching} captures the mean absolute difference of consecutive resolutions; lower values indicate smoother visual transitions.
\end{enumerate*}
%%%%%%%%%%
\subsection{Methods used for comparisons}
We compare \DQL against four baseline methods:
\begin{enumerate*}
\item\emph{HLS~\cite{HLS_ladder_ref}} uses a fixed bitrate-resolution ladder defined by Apple's specification (Table~\ref{tab:hls}).
\item\emph{CDBL~\cite{azimi2024decoding}}selects the encoding configuration with the highest XPSNR, satisfying a given decoding time constraint and ensuring the decoding time remains below a threshold $\tau_L$.
\item\emph{RQT-PF~\cite{katsenou2024decoding}}  minimizes a composite cost function, balancing decoding time and bitrate, controlled by a trade-off parameter $\alpha$, without directly optimizing video quality metrics such as XPSNR or VMAF.
\item\emph{VEXUS~\cite{menon2024convex}} optimizes the bitrate ladder for bitrate-XPSNR trade-offs, maximizing quality and minimizing bitrate without considering decoding complexity.
\end{enumerate*}
%%%%%%%%%%
\section{Evaluation Results}
\label{sec:EResult} 
This section evaluates prediction models, DQN agent behavior, and bitrate ladder efficiency compared to four baselines.
%%%%%%%%%
\begin{table*}[!t]
    \centering
    \begin{minipage}{0.49\linewidth}
        \centering
        \caption{Bitrate prediction results.}
        \label{tab:average_results_bitrate}
        \fontsize{7pt}{7pt}\selectfont
        \begin{tabular}{|c|c|c|c|c|c|}
            \hline
            \textit{Model} & $R^2$ & \textit{RMSE} & \textit{SDAE} & \textit{MAE\%} & \textit{Inf. Time} \\ \hline
            \textbf{\textit{XGBoost}}    & \textbf{0.89} & \textbf{4651.06}  & \textbf{4513.09}  & \textbf{11.17} & \textbf{0.6} \\ \hline
            \textit{LightGBM}   & 0.89 & 6971.99  & 6489.9   & 24.8  & 0.53 \\ \hline
            \textit{LR}         & 0.82 & 8816.12  & 8156.82  & 32.6  & 0.003 \\ \hline
            \textit{RF}         & 0.87 & 7390.21  & 6862.91  & 26.71 & 1.84 \\ \hline
            \textit{ExtraTrees} & 0.86 & 7722.34  & 7185.67  & 27.56 & 0.32 \\ \hline
            \textit{Adaboost}   & 0.71 & 11295.59 & 10617.74 & 37.55 & 0.32  \\ \hline
           \textit{MLP}        & 0.91 & 6091.83  & 5631.26  & 22.84 & 0.8 \\ \hline
        \end{tabular}
    \end{minipage}
    \hfill
    \begin{minipage}{0.49\linewidth}
\centering
\caption{Decoding time prediction results.}
\label{tab:average_results_dec}
\fontsize{7pt}{7pt}\selectfont
\begin{tabular}{|c|c|c|c|c|c|}
\hline
\textit{Model} & \textit{$R^2$} & \textit{RMSE} & \textit{SDAE} & \textit{MAE\%} & \textit{Inference time} \\ \hline
\textit{XGBoost}    & 0.94 & 4.48 & 3.96 & 12.46 & 0.5 \\ \hline
\textit{LightGBM}   & 0.95 & 4.2  & 3.7  & 12.02 & 0.56 \\ \hline
\textit{LR}         & 0.75 & 9.43 & 7.14 & 36.8  & 0.002 \\ \hline
\textbf{\textit{RF}}  & \textbf{0.96} & \textbf{3.9}  & \textbf{3.41} & \textbf{11.49} & \textbf{0.31} \\ \hline
\textit{ExtraTrees} & 0.94 & 4.58 & 4.02 & 13.04 & 0.08 \\ \hline
\textit{Adaboost}   & 0.85 & 7.37 & 4.08 & 36.64 & 0.33 \\ \hline
\textit{MLP}         & 0.96 & 3.87 & 3.3  & 12.05 & 0.7 \\ \hline
\end{tabular}
\end{minipage}
    \vspace{3mm}

    \begin{minipage}{0.49\linewidth}
        \centering
        \caption{XPSNR prediction results.}
        \label{tab:average_results_xpsnr}
        \fontsize{7pt}{7pt}\selectfont
        \begin{tabular}{|c|c|c|c|c|c|}
            \hline
             \textit{Model} & $R^2$ &  \textit{RMSE} &  \textit{SDAE} &  \textit{MAE\%} &  \textit{Inf. Time} \\ \hline
             \textbf{\textit{XGBoost}}    & \textbf{0.96} & \textbf{1.22} & \textbf{0.79}  & \textbf{2.51} & \textbf{0.44} \\ \hline
             \textit{LightGBM}   & 0.96 & 1.23 & 0.79  & 2.53 & 0.56 \\ \hline
             \textit{LR}         & 0.86 & 2.2  & 1.4   & 4.53 & 0.002 \\ \hline
             \textit{RF}         & 0.94 & 1.42 & 0.9   & 2.97 & 1.55 \\ \hline
             \textit{ExtraTrees} & 0.94 & 1.45 & 0.95  & 2.94 & 0.34 \\ \hline
             \textit{Adaboost}   & 0.92 & 1.62 & 1.001 & 3.42 & 0.32 \\ \hline
             \textit{MLP}        & 0.96 & 1.16 & 0.74  & 2.39 & 0.92 \\ \hline
        \end{tabular}
    \end{minipage}
    \hfill
    \begin{minipage}{0.49\linewidth}
        \centering
        \caption{VMAF prediction results.}
        \label{tab:average_results_vmaf}
        \fontsize{7pt}{7pt}\selectfont
        \begin{tabular}{|c|c|c|c|c|c|}
            \hline
             \textit{Model} & $R^2$ &  \textit{RMSE} &  \textit{SDAE} &  \textit{MAE\%} &  \textit{Inf. Time} \\ \hline
             \textbf{\textit{XGBoost}}    & \textbf{0.93} & \textbf{6.38}  & \textbf{4.44} & \textbf{6.74}  & \textbf{0.42} \\ \hline
             \textit{LightGBM}   & 0.93 & 6.45  & 4.52 & 6.77  & 0.56 \\ \hline
            LR         & 0.75 & 11.74 & 6.75 & 14.13 & 0.002 \\ \hline
             \textit{RF}         & 0.93 & 6.22  & 4.33 & 6.57  & 1.57 \\ \hline
             \textit{ExtraTrees} & 0.92 & 6.82  & 4.76 & 7.17  & 0.34 \\ \hline
             \textit{Adaboost}   & 0.85 & 9.22  & 4.6  & 11.75 & 0.32 \\ \hline
             \textit{MLP}        & 0.94 & 5.99  & 3.69 & 6.92  & 0.84 \\ \hline
        \end{tabular}
    \end{minipage}
\end{table*}
%%%%
%%
\begin{figure*}[!t]
    \centering \includegraphics[width=.9\linewidth]{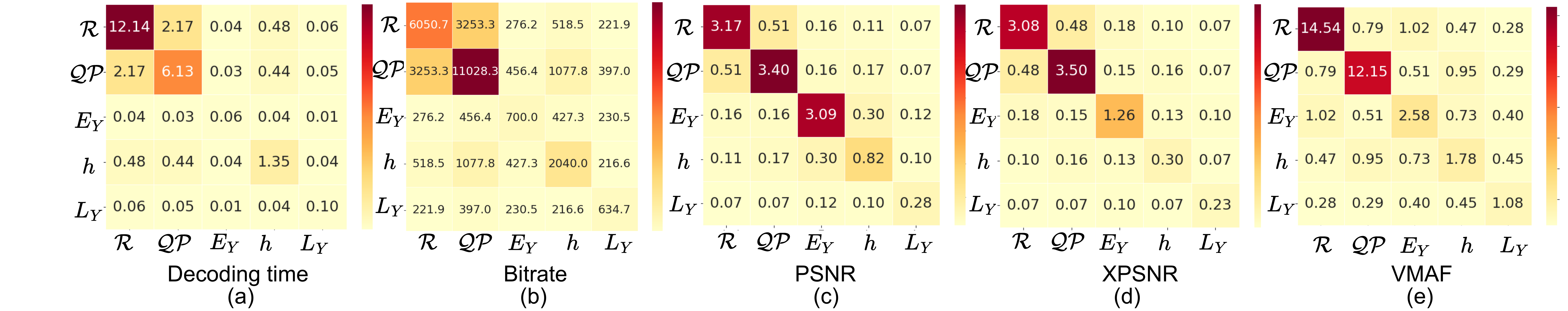}
    \caption{SHAP analysis of feature contribution and interaction across prediction metrics.}
    \label{fig:shap}
\end{figure*}
%%%
\subsection{\DQL prediction models analysis}\label{prediction-results}
We evaluated the performance of candidate prediction models on four target metrics: decoding time, bitrate, XPSNR, and VMAF. The reported results represent the average performance across the entire test set; however, the inference time applies only per segment during online ladder construction, making its overhead negligible. Each target metric shows distinct characteristics influencing model suitability. For example, decoding time shows regular patterns favoring tree-based models like RF. Meanwhile, bitrate and quality scores (i.e., VMAF, XPSNR) exhibit more complex, non-linear dependencies on encoding parameters, where gradient boosting and neural-based models (MLP) demonstrate higher accuracy. We assess models using the evaluation metrics in Section~\ref{metric-predic} and select the best-performing model for each target by considering prediction accuracy and inference efficiency.
\subsubsection{Decoding time prediction} 
Table~\ref{tab:average_results_dec} shows that RF achieved the highest $R^2$ (\num{0.96}) and the lowest RMSE (\num{3.9}), closely matching MLP performance. However, RF had a lower MAE\% (\num{11.49}) and faster inference time (\qty{0.31}{\second} vs. \qty{0.7}{\second}), becoming the preferred model for decoding time prediction. The best RF performance was with $n_{trees} = 100$ and $d_{\max} = 10$.
%%%%
\subsubsection{Bitrate prediction} Table~\ref{tab:average_results_bitrate} shows that  MLP achieved the highest $R^2$ (\num{0.91}), while XGBoost delivered better absolute error metrics (RMSE \num{4651} vs. \num{6091}, MAE\% \num{11.17} vs. \num{22.84}) and lower inference time (\qty{0.6}{\second} vs. \qty{0.8}{\second}). Thus, we selected XGBoost for bitrate prediction because of its better trade-off between accuracy and efficiency. XGBoost achieved its best performance with $n_{trees} = 300$, $d_{\max} = 7$, and $\eta = 0.5$.
\subsubsection{XPSNR prediction} 
Table~\ref{tab:average_results_xpsnr} shows that the MLP once again achieved the lowest RMSE (\num{1.16}) and MAE\% (\num{2.39}). Although XGBoost and LightGBM delivered performance comparable to MLP, XGBoost’s lower inference time (\qty{0.44}{\second} vs. \qty{0.56}{\second} and \qty{0.92}{\second}) makes it the preferred model. The XGBoost model achieved its best performance with $n_{trees} = 300$, $d_{\max} = 7$, and $\eta = 0.1$.
\subsubsection{VMAF prediction} 
As shown in Table~\ref{tab:average_results_vmaf}, the MLP achieved the best results across all metrics, with an $R^2$ of \num{0.94}, RMSE of \num{5.99}, and MAE\% of \num{6.92}. Similarly to the XPSNR case, XGBoost and LightGBM exhibited performance close to MLP. We chose XGBoost for VMAF prediction due to its lower inference time (\qty{0.42}{\second}). XGBoost achieved its best performance with $n_{trees} = 300$, $d_{\max} = 5$, and $\eta = 0.5$.
%%%
\subsubsection{Feature importance analysis}
%%%
\begin{figure}[!t]
    \centering    
    \includegraphics[width=.85\linewidth]{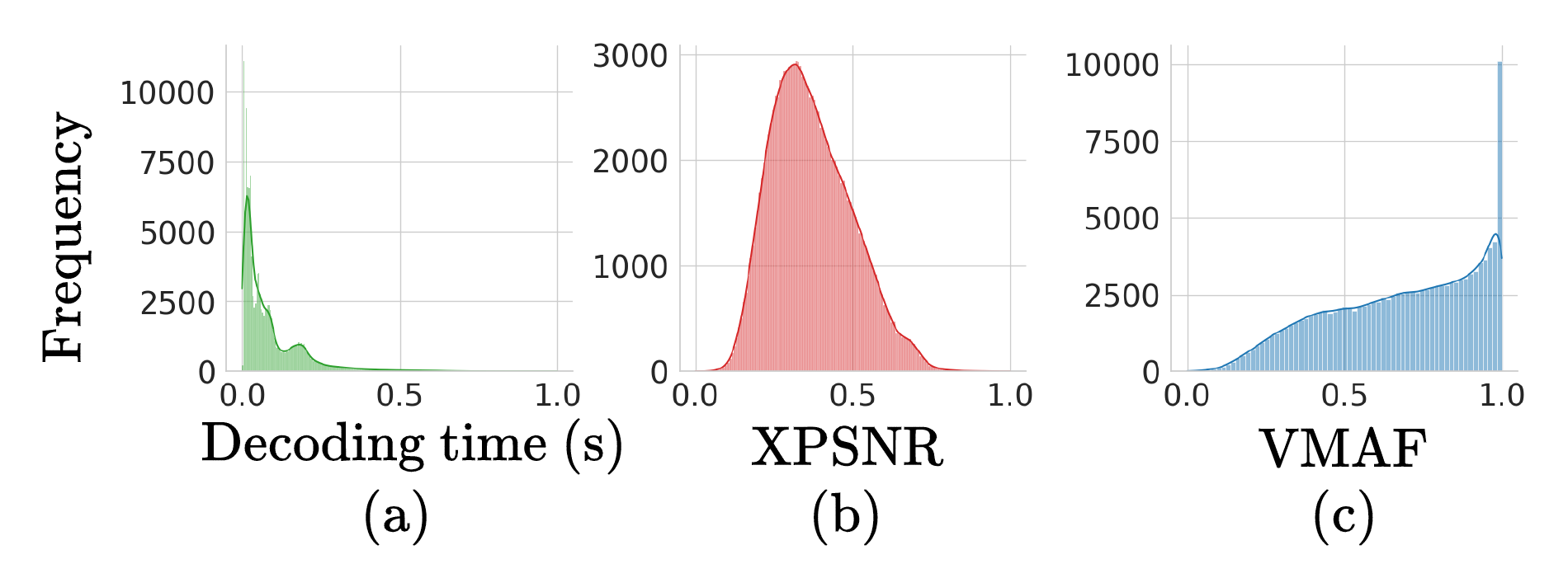}
    \caption{Distribution of normalized decoding time and quality metrics (XPSNR, VMAF).}
    \label{Weighting-dist}
\end{figure}
%%%
We employed \emph{SHapley Additive exPlanations} (SHAP)~\cite{lundberg2017unified} to show the contribution of each input feature to the prediction targets. SHAP values capture the average impact of a feature on model outputs, while SHAP interaction values reveal dependencies between feature pairs. Fig.~\ref{fig:shap} visualizes the mean absolute SHAP values and interaction heatmaps.
Across all prediction metrics, quantization parameters ($\mathcal{QP}$) and resolutions ($\mathcal{R}$) have the dominant features influencing model predictions. Spatial complexity ($E_Y$) also plays a key role across both quality metrics. In contrast, temporal complexity ($h$) has a higher impact on bitrate and decoding time due to its relation to motion cost. Moreover, the strongest feature interaction is observed between $\mathcal{QP}$ and $\mathcal{R}$, particularly for decoding time and bitrate prediction, reflecting their impact on compression and computational complexity. Luminance variation ($L_Y$) shows the least contribution across all metrics.
%%%%%%%%%%%%%%%%%%%%%%%%%%%%%%
\subsection{\DQL agent analysis}\label{DQN_result}
We analyze the DQN agent’s behavior using four experiments described in this section.
%%%%%%%%%%%%%
\subsubsection{Reward normalization and weighting strategy} \label{distribution} We analyzed the normalized distributions of decoding time and quality metrics to guide the reward function (Eq.~\ref{rew}) precisely. As shown in Fig.~\ref{Weighting-dist}, even after normalization to $[0,1]$, the distributions differ significantly: decoding time is heavily right-skewed (skewness=\num{2.4}), XPSNR is moderately right-skewed (\num{0.49}), and VMAF is slightly left-skewed (\qty{-0.42}), indicating that equal weights $\lambda_1$ and $\lambda_2$ cannot balance contributions in the reward function. Thus, we relax the constraint \mbox{$\lambda_1 + \lambda_2 + \lambda_3 = 1$} and allow independent tuning, enabling more flexible and data-driven trade-offs between quality and decoding efficiency.
%%%%%%%%%%%%%
\subsubsection{Cumulative reward and convergence analysis}
\begin{figure}[!t]
    \centering    
    \includegraphics[width=0.8\linewidth]{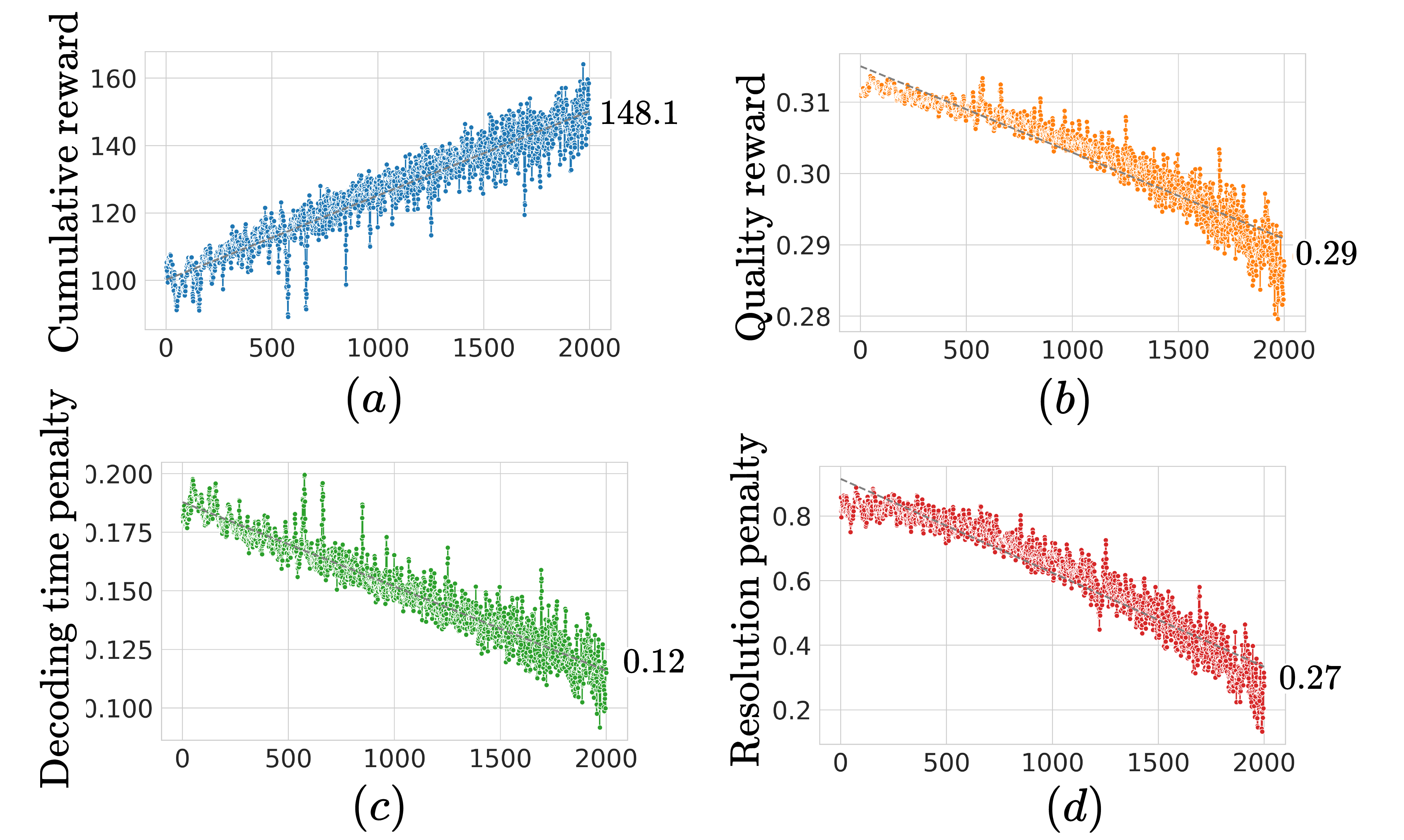}
    \caption{Convergence trends of the DQN agent for (a) cumulative reward, (b) XPSNR-based quality reward, (c) decoding time penalty, and (d) resolution smoothness penalty.}
    \label{fig:conv}
\end{figure}

We monitored \num{2000} training episodes to assess the DQN agent’s convergence behavior. As shown in Fig.~\ref{fig:conv} (a), the cumulative reward increased steadily, confirming progressive policy improvement. The XPSNR reward Fig.~\ref{fig:conv} (b) declined slightly by \qty{7.4}{\percent}, indicating the agent’s strategic shift toward reducing time complexity. Decoding time and resolution smoothness penalties dropped by \qty{36.1}{\percent} (Fig.~\ref{fig:conv} (c)) and \qty{65.8}{\percent} (Fig.~\ref{fig:conv} (d)), respectively, showing that the agent learned to balance quality with efficiency.
%%%
%%%%%%%%%%%%%%%%
%%%
\begin{figure}[!t]
    \centering    
    \includegraphics[width=.8\linewidth]{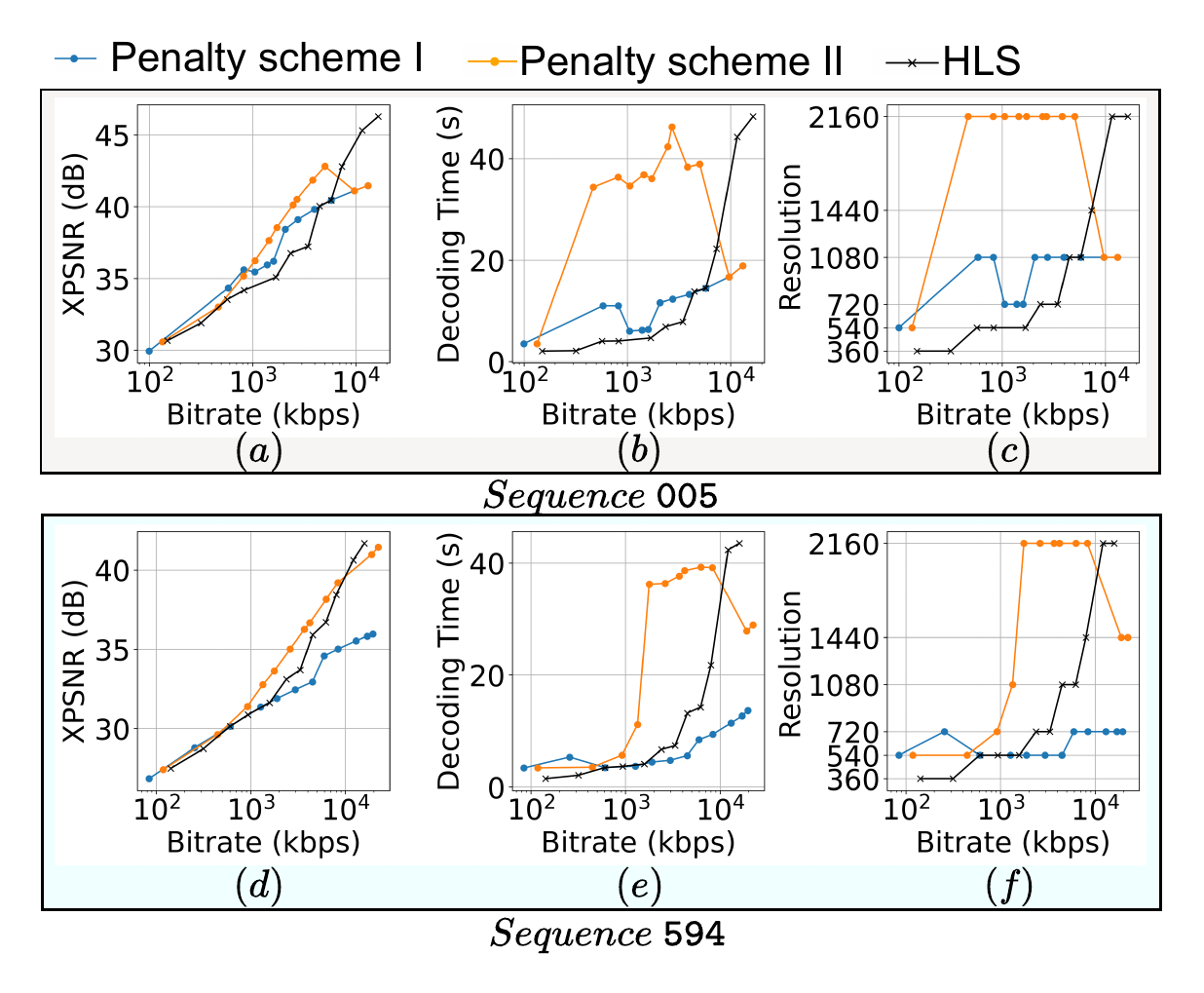}
    \caption{Impact of different penalties on \DQL, evaluated on (a–c) sequences \texttt{005} and (d–e) \texttt{594}.}
    \label{fig:res_imp}
\end{figure}
%%%
\begin{figure*}
    \centering    
    \includegraphics[width=.8\linewidth]{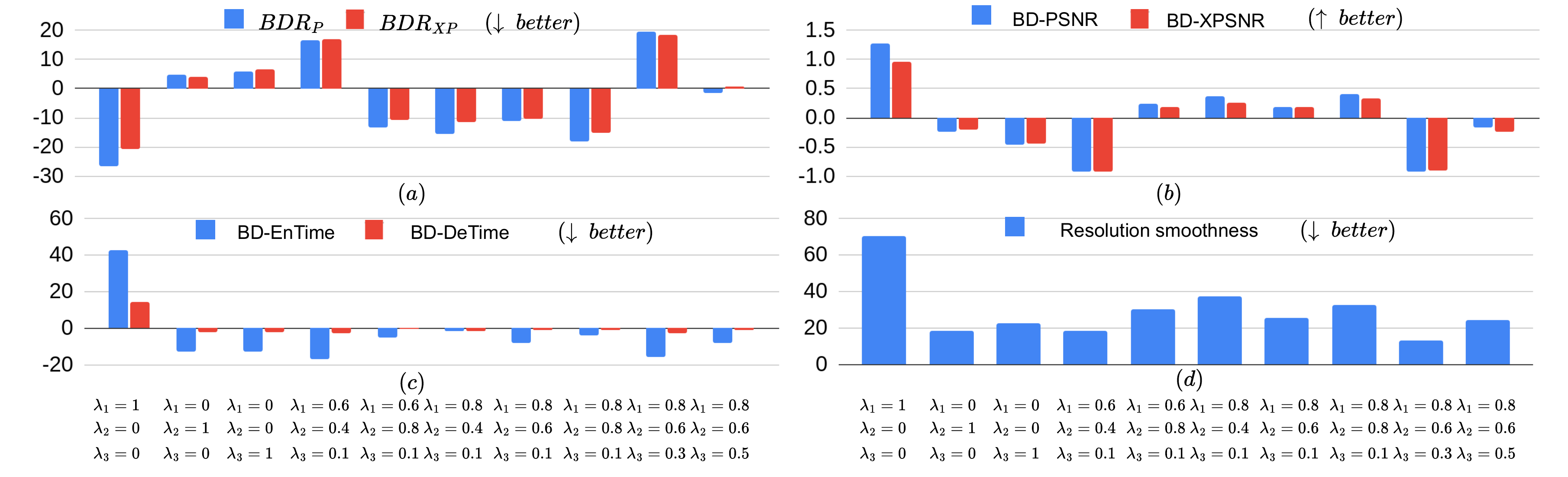}
    \caption{Impact of different $\lambda_1$, $\lambda_2$, and $\lambda_3$ values on \DQL performance outcomes.}
    \label{fig:bar_dql}
\end{figure*}
%%%%
\subsubsection{Penalty impact} We evaluated two resolution-switching penalty strategies reflecting provider-specific example policies: \textit{(i)} a fixed penalty applied whenever the selected resolution differs from the previous one (Eq.~\ref{eq:penalty}), uniformly discouraging changes and favoring stable transitions; \textit{(ii)} a history-aware penalty, triggered when the new resolution deviates from the two most recent selections, scaled by the normalized resolution difference, promoting flexible yet controlled adaptation. To isolate their effects, we set \mbox{$\lambda_1=0,\lambda_2=0,\lambda_3=1$}, enforcing resolution-switch minimization as the sole optimization objective. Fig.~\ref{fig:res_imp} (c) and (f) show that the fixed strategy (blue line) yields a conservative ladder with minimal switching, while the history-aware variant (orange line) allows smoother resolution rises. These schemes reveal inherent trade-offs between stability and adaptability, allowing providers to tailor switching behavior to application-specific requirements.
%%%%
\subsubsection{System overhead} We analyzed the end-to-end latency introduced by \DQL to assess its suitability for scalable scenarios. Table~\ref{tab:DQL-Delay} reports the average delay per representation (rep.) and \qty{4}{\second} segment (seg.), including three main components: VCA, metric predictions, and DQN agent decision. Since prediction models operate in parallel, the maximum inference time among the three predictors can be considered. The results show that \DQL imposes negligible delay per representation (sub-millisecond), and sub-second latency for a full ladder construction per segment, confirming its feasibility for real-time bitrate selection. 
%%%%%
\begin{table}[!t]
\centering 
\caption{Average delay imposed by \DQL.} 
\label{tab:DQL-Delay}
\fontsize{7pt}{7pt}\selectfont
\begin{tabular}{|c|c|c|c|c|c|c|c|}
\hline
\multirow{3}{*}{\textit{Granularity}} &
\multirow{3}{*}{\textit{\begin{tabular}[c]{@{}c@{}}VCA \\ (s)\end{tabular}}} &
\multicolumn{4}{c|}{\textit{Metrics prediction (ms)}} &
\multirow{3}{*}{\textit{\begin{tabular}[c]{@{}c@{}}DQN \\ agent \\ (ms)\end{tabular}}} &
\multirow{3}{*}{\textit{\begin{tabular}[c]{@{}c@{}}Total \\ (s)\end{tabular}}} \\ \cline{3-6}
& &
\multicolumn{2}{c|}{\textit{Quality}} &
\multirow{2}{*}{\textit{Bitrate}} &
\multirow{2}{*}{\textit{\begin{tabular}[c]{@{}c@{}}Dec. \\ time\end{tabular}}} &
& \\ \cline{3-4}
& &
\textit{XPSNR} &
\textit{VMAF} &
& &
& \\ \hline
\textit{Rep.} &
\multirow{2}{*}{2.71} &
0.12 &
0.23 &
0.16 &
0.16 &
0.3 &
2.71 \\ \cline{1-1} \cline{3-8}
\textit{Seg.} &
&
1.4 &
2.7 &
1.95 &
1.87 &
3.6 &
2.72 \\ \hline
\end{tabular}
\end{table}
%%%%%%%%
%%%%%
\begin{figure*}[!t]
    \centering    
    \includegraphics[width=1\linewidth]{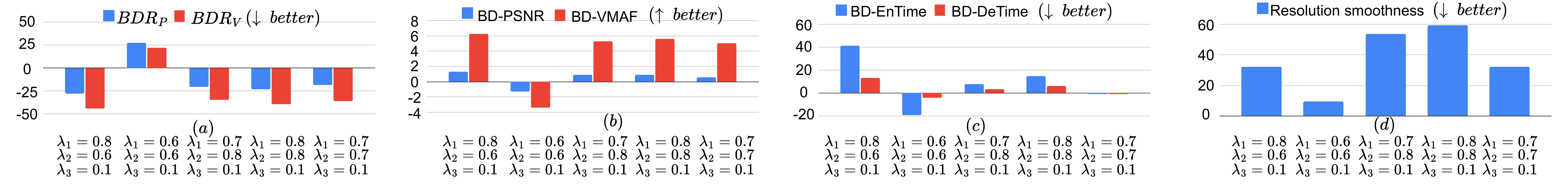}
    \caption{Impact of varying $\lambda_1$, $\lambda_2$, and $\lambda_3$ values on \DQL performance for the VMAF metric.}
    \label{fig:vmaf}
\end{figure*}
%%%%%
\subsection{\DQL weighting coefficients analysis}\label{DQN_result_weight}
We analyze how different weighting coefficients influence \DQL behavior through the following experiments.
\subsubsection{Impact on bitrate ladder performance}
We changed the \mbox{$\lambda_1$--$\lambda_3$} values to analyze the impact of different reward weightings. As shown in Fig.~\ref{fig:bar_dql}, setting \mbox{(1,0,0)} maximizes quality (Fig.~\ref{fig:bar_dql} (a)), improving PSNR and XPSNR by \num{1.26} dB and \num{0.96} dB over HLS (Fig.~\ref{fig:bar_dql} (b)), but at the cost of a \qty{42.7}{\second} encoding and \qty{14.4}{\second} decoding time increase (Fig.~\ref{fig:bar_dql} (c)). In contrast, with ($0,1,0$), encoding and decoding times are reduced by \qty{12.9}{\second} and \qty{2.1}{\second}, respectively, but incur approximately \num{2} dB quality loss and unstable resolution shifts (Fig.~\ref{fig:bar_dql} (d)). To balance these trade-offs, we evaluated other configurations. Among them, \mbox{($0.8,0.6,0.1$)} offers the best trade-off: a BD-rate reduction of \qty{10.94}{\percent} (PSNR) and \qty{10.3}{\percent} (XPSNR), encoding and decoding time savings of \qty{7.96}{\second} and \qty{0.72}{\second}, and improved resolution smoothness, all without degrading quality.
%%%%%%%%%%%
\subsubsection{Impact on VMAF}
Starting from the initial weight configuration optimized for XPSNR, we changed the values of \mbox{$\lambda_1$--$\lambda_3$} to analyze the impact of different reward weightings on VMAF improvement. In Fig.~\ref{fig:vmaf}~(a), configurations with higher $\lambda_1$ and $\lambda_2$, such as $(0.7,0.8,0.1)$ and $(0.8,0.8,0.1)$, achieve the lowest $BDR_V$, indicating better bitrate efficiency for VMAF. Fig.~\ref{fig:vmaf}~(b) shows that BD-VMAF improves by approximately 5--6\% with these settings, while BD-PSNR is almost identical, confirming quality gains. In terms of complexity (Fig.~\ref{fig:vmaf}~(c)), $(0.7,0.8,0.1)$ reduces encoding and decoding times compared to $(0.8,0.8,0.1)$, maintaining efficiency. In addition, resolution smoothness (Fig.~\ref{fig:vmaf}~(d)) is also improved, with $(0.7,0.8,0.1)$ minimizing resolution switching artifacts. Based on these results, we select \mbox{$(\lambda_1=0.7, \lambda_2=0.7, \lambda_3=0.1)$} as the weighting configuration,  yielding a BD-VMAF gain of \num{5} points, \qty{36}{\percent} bitrate reduction, a BD-DeTime of \qty{-0.71}, and smooth resolution transitions, without compromising quality compared to HLS.
%%%%%%%%%%%
\begin{figure*}[!t]
    \centering    
    \includegraphics[width=.9\linewidth]{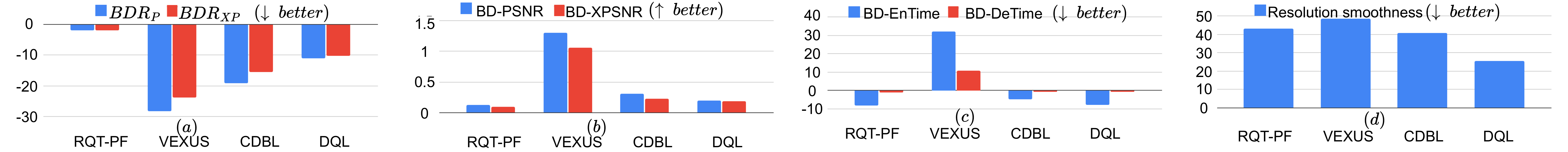}
    \caption{Comparison of \DQL and baseline schemes in terms of quality, time, and switching smoothness.}
    \label{fig:bar_sota}
\end{figure*}
%%%%%
\subsection{\DQL vs. baseline analysis}
We evaluated the performance of \DQL under the selected configuration \mbox{($\lambda_1=0.8, \lambda_2=0.6, \lambda_3=0.1$)} against four state-of-the-art bitrate ladder construction schemes. 
%%%
\subsubsection{Impact on bitrate ladder performance}
Fig.~\ref{fig:bar_sota} compares the average performance of \DQL against RQT-PF (with $\alpha=0.75$), CDBL (with $\tau_L=16$), and VEXUS. VEXUS yields the highest BD-PSNR (\num{+1.3} dB) and BD-XPSNR (+\num{1.05} dB), but at the cost of increased encoding time (\qty{+32.2}{\second}) and decoding time (\qty{+10.7}{\second}). \DQL achieves a balance with BD-rate of \qty{-10.94}{\percent} (PSNR) and \qty{-10.3}{\percent} (XPSNR), while improving BD-PSNR by \num{0.19} dB and BD-XPSNR by \num{0.18} dB.
\DQL reduces bitrate overhead and improves quality, while maintaining almost similar encoding and decoding durations compared to RQT-PF. \DQL also saves \qty{6.1}{\second} in encoding time and \qty{0.3}{\second} in decoding time, at a marginal bitrate cost compared to CDBL.
Moreover, \DQL achieves the lowest resolution smoothness score (\num{25.5}), having fewer resolution switches than all baselines.
%%%

\subsubsection{Impact on robustness to prediction errors}
\begin{figure}[!t]
    \centering    
    \includegraphics[width=.95\linewidth]{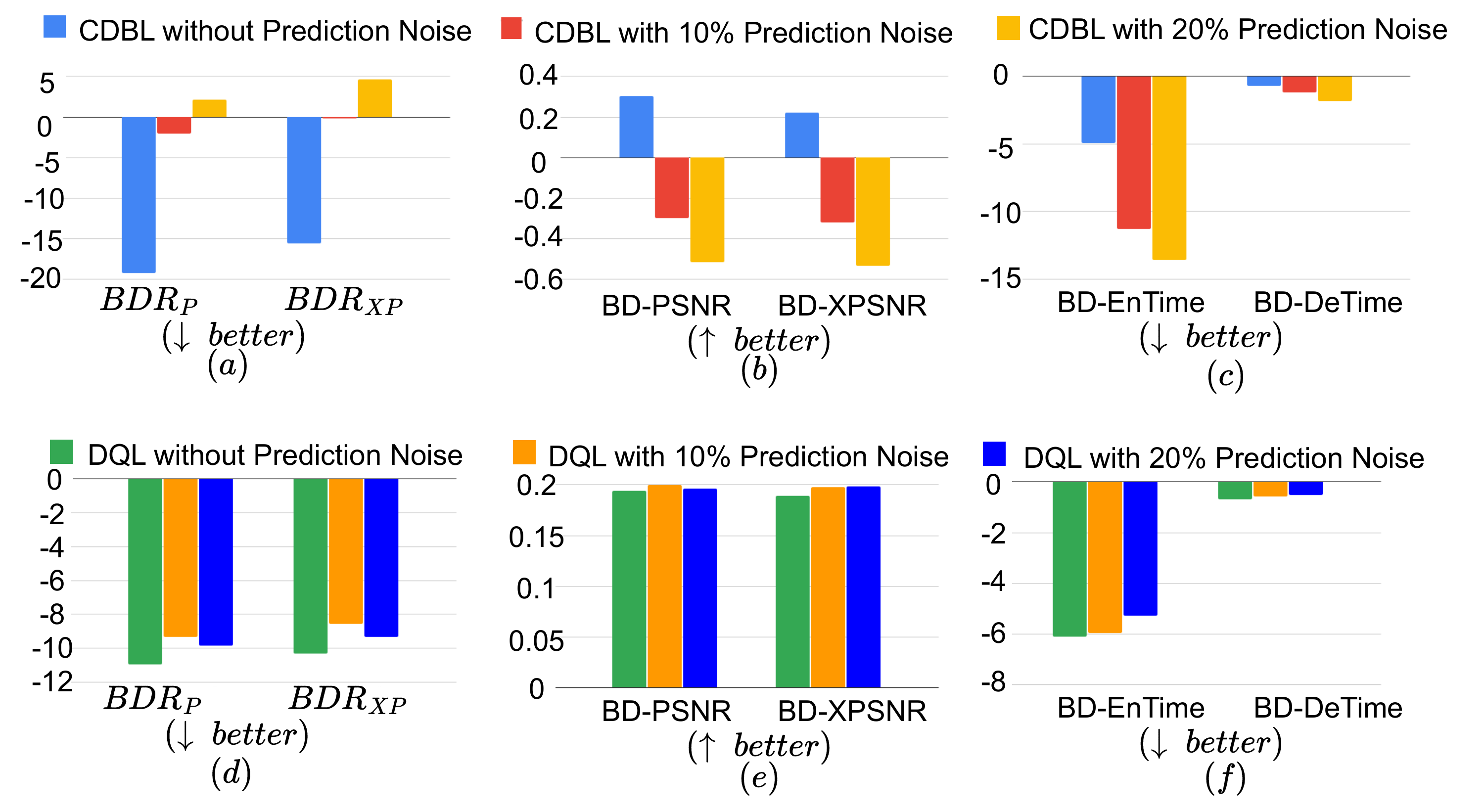}
    \caption{Robustness of CDBL and \DQL under increasing prediction noise.}
    \label{fig:noise}
\end{figure}

Threshold-based optimization methods such as CDBL derive ladder decisions directly from predicted decoding time and quality values, making them sensitive to estimation noise when predictions cross decision boundaries. To quantify this effect, we inject additive Gaussian noise of up to \qty{20}{\percent} into the predicted decoding time and XPSNR metrics, following established evaluation practice~\cite{zhang2022bilateral, wisniewski2025benchmarking}. As shown in Fig.~\ref{fig:noise}(a–c), CDBL shows pronounced instability under noisy predictions, where its BD-rate degrades from \qty{-19.26}{\percent} under base predictions to \qty{-2.04}{\percent} at \qty{10}{\percent} noise and becomes positive (\qty{+2.10}{\percent}) at \qty{20}{\percent} noise. Moreover, its BD-PSNR and BD-XPSNR simultaneously drop below zero, indicating loss of both compression efficiency and quality gains. In contrast, \DQL remains robust across all noise levels (Fig.~\ref{fig:noise}(d–f)). Even with \qty{20}{\percent} noise, \DQL BD-rate and BD-quality metrics remain close to the baseline setting, with only minor increases in encoding and decoding time.
%%%%%
%%%%%
\subsubsection{Impact on video sequences}
\begin{figure}[!t]
    \centering    
    \includegraphics[width=.9\linewidth]{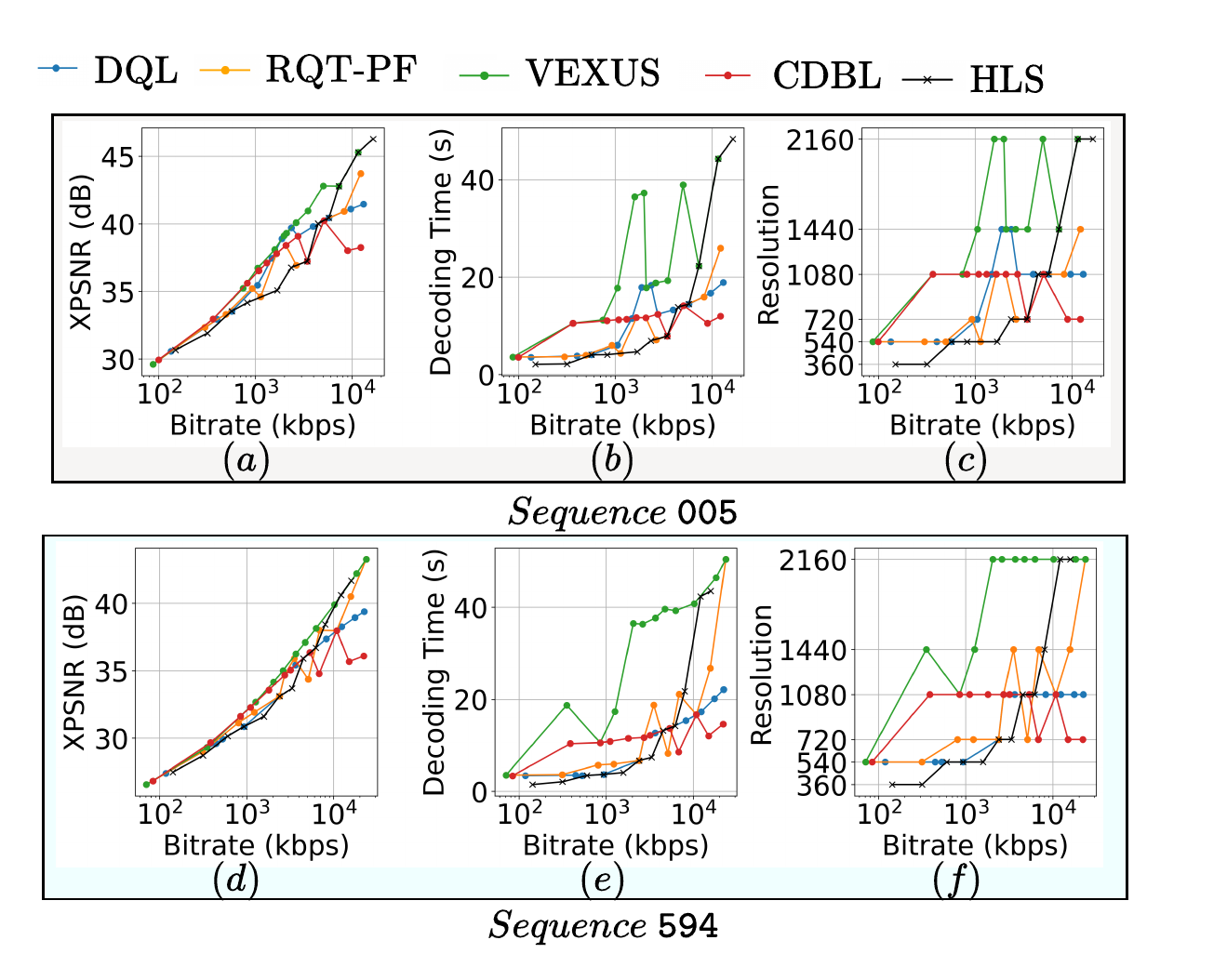}
    \caption{\DQL vs. RQT-PF, VEXUS, and CDBL on video sequence \texttt{005} (a-c), \texttt{594}(d-f).}
    \label{fig:per_segment_sota}
\end{figure}

Fig.~\ref{fig:per_segment_sota} illustrates bitrate ladders for two representative video segments with distinct complexity characteristics. VEXUS achieves the highest XPSNR across both segments, albeit with higher decoding time. The resolution smoothness penalty in \DQL leads to more stable trends in both XPSNR and decoding time across the ladder. In contrast, RQT-PF and CDBL exhibit frequent resolution switches, resulting in noticeable fluctuations in both metrics. Consequently, the \DQL-based ladder shows a more consistent progression, maintaining high XPSNR while keeping decoding time low.
%%%%%
%%%%%%%%%%%%%%%%%%%%%%%%%%%%%%%%%%%
\section{Conclusion}
\label{sec:conclusion} 
This article presented \DQL, a deep Q-learning (DQL)–based framework for constructing adaptive bitrate ladders by jointly considering video quality, decoding time, and resolution stability.
We developed a two-phase design for \DQL: 1) a training phase, in which the agent is optimized using real encoding feedback and a multi-objective reward that balances video quality, decoding time, and resolution stability; and 2) an inference phase, where lightweight content features and predictive models enable efficient bitrate ladder construction without invoking real encoders. Extensive evaluation on \num{750} diverse real-world video sequences shows that \DQL outperforms state-of-the-art methods. Future work focuses on extending \DQL to incorporate network dynamics and client-side feedback for end-to-end adaptive streaming optimization.
\bibliographystyle{./bibliography/IEEEtran}
\bibliography{./bibliography/IEEEabrv}

@inproceedings{psnr_ref,
  title={{Objective Video Quality Measurement Using A Peak-Signal-to-Noise Ratio Full Reference Technique}},
  author={{Alliance For Telecommunications Industry Solutions}},
  booktitle={T1.TR.74-2001},
  year={2001}
}

@article{ssim_ref,
  title={{Image Quality Assessment: From Error Visibility to Structural Similarity}},
  author={Wang, Zhou and others},
  journal={{IEEE Trans. on Image Processing}},
  year={2004},
  publisher={{IEEE}}
}

@INPROCEEDINGS{itu_xpsnr_vs_vmaf,
author={Helmrich, Christian R. and others},
booktitle={ITU Journal: ICT Discoveries},
year={2020},
title={{A Study of the Extended Perceptually Weighted Peak Signal-to-Noise Ratio for Video Compression with Different Resolutions and Bit Depths}}
}

@INPROCEEDINGS{wien_xpsnr_vs_vmaf,
author={Wien, Mathias and Baroncini, V.},
title={{Report on VVC Compression Performance Verification Testing in the SDR UHD Random Access Category}},
year={2020},
booktitle={WG 05 MPEG Joint Video Coding Team(s) with ITU-T SG 16, Document JVET-T0097},
}

@INPROCEEDINGS{xpsnr_vs_vmaf,
author={{C. R. Helmrich} and others},
title={{Information On and Analysis of the VVC Encoders in the SDR UHD Verification Test}},
year={2020},
booktitle={WG 05 MPEG Joint Video Coding Team(s) with ITU-T SG 16, Document JVET-T0103},
}

@article{inter4k_ref,
  title={{Adapool: Exponential Adaptive Pooling for Information-retaining Downsampling}},
  author={Stergiou, Alexandros and Poppe, Ronald},
  journal={{IEEE Trans. on Image Processing}},
  year={2022},
  publisher={{IEEE}}
}

@inproceedings{vvenc_ref,
  title={{VVenC: An Open and Optimized VVC Encoder Implementation}},
  author={Wieckowski, Adam and others},
  booktitle={{IEEE Intl. Conf. on Multimedia \& Expo Workshops}},
  year={2021},
  organization={{IEEE}}
}

@inproceedings{DASH_ref,
  title={{Dynamic Adaptive Streaming Over HTTP-- Standards and Design Principles}},
  author={Stockhammer, Thomas},
  booktitle={{Proc. of the Second Annual ACM Conf. on Multimedia Systems}},
  year={2011}
}

@inproceedings{gnostic,
  title={{Content-gnostic Bitrate Ladder Prediction for Adaptive Video Streaming}},
  author={Katsenou, Angeliki V and others},
  booktitle={{Picture Coding Symposium}},
  year={2019},
  organization={{IEEE}}
}

@article{jnd_ref,
  title={{Visual JND: A Perceptual Measurement in Video Coding}},
  author={Yuan, Di and others},
  journal={{IEEE Access}},
  year={2019},
  publisher={{IEEE}}
}

@book{lr_ref,
  title={{Introduction to Linear Regression Analysis}},
  author={Montgomery, Douglas C and others},
  year={2021},
  publisher={{John Wiley \& Sons}}
}

@article{rf_ref,
  title={{Random Forests}},
  author={Breiman, Leo},
  journal={{Machine Learning}},
  year={2001},
  publisher={{Springer}}
}

@inproceedings{xgboost_ref,
  title={{XGBoost}: {A} {Scalable} {Tree} {Boosting} {System}},
  author={Chen, Tianqi and Guestrin, Carlos},
  booktitle={Proc. of the 22nd {ACM} {SIGKDD} {Intl.} {Conf.} on {Knowledge} {Discovery} and {Data} {Mining}},
  year={2016}
}

@inproceedings{vca_ref,
  title={{Green Video Complexity Analysis for Efficient Encoding in Adaptive Video Streaming}},
  author={Menon, Vignesh V and others},
  booktitle={{Proc. of the First Intl. Workshop on Green Multimedia Systems}},
  year={2023}
}

@article{kingma2014adam,
  title={{Adam: A Method for Stochastic Optimization}},
  author={Kingma, Diederik P and Ba, Jimmy},
  journal={arXiv preprint arXiv:1412.6980},
  year={2014}
}

@INPROCEEDINGS{VMAF,
  title={{VMAF: The Journey Continues}},
  author={Li, Zhi and others},
  booktitle={{Netflix Technology Blog}},
  year={2018}
}

@misc{grid, 
   title = {{GridSearchCV}},
    url = {https://scikit-learn.org/stable/modules/generated/sklearn.model_selection.GridSearchCV.html},
  note = {{Retrieved: 2026-02-06}}
  }

@inproceedings{hameed2016decision,
  title={{A Decision-tree-based Perceptual Video Quality Prediction Model and its Application in FEC for Wireless Multimedia Communications}},
  author={Hameed, Abdul and others},
  booktitle={{IEEE Trans. on Multimedia}},
  year={2016},
  publisher={{IEEE}}
}

@article{ickin2021qoe,
  title={{QoE Modeling on Split Features with Distributed Deep Learning}},
  author={Ickin, Selim and others},
  journal={{Network}},  
  year={2021},
  publisher={{MDPI}}
}

@inproceedings{begluk2018machine,
  title={{Machine Learning-based QoE Prediction for Video Streaming Over LTE Network}},
  author={Begluk, Tarik and others},
  booktitle={{17th Intl. Symp. INFOTEH-JAHORINA}},
  year={2018},
  organization={{IEEE}}
}

@article{menon2023jnd,
  title={{JND-Aware Two-Pass Per-Title Encoding Scheme for Adaptive Live Streaming}},
  author={Menon, Vignesh V and others},
  journal={{IEEE Trans. on Circuits and Systems for Video Technology}},
  year={2023},
  publisher={{IEEE}}
}

@article{lebreton2023quitting,
  title={{Quitting Ratio-based Bitrate Ladder Selection Mechanism for Adaptive Bitrate Video Streaming}},
  author={Lebreton, Pierre and Yamagishi, Kazuhisa},
  journal={{IEEE Trans. on Multimedia}},
  year={2023},
  publisher={{IEEE}}
}

@article{zou2017event,
  title={{Event-based Perceptual Quality Assessment for HTTP-based Video Streaming with Playback Interruption}},
  author={Zou, Wenjie and others},
  journal={{IEEE Trans. on Multimedia}},
  year={2017},
  publisher={{IEEE}}
}

@inproceedings{res_pred_ref1,
  title={{Combining Video Quality Metrics to Select Perceptually Accurate Resolution in a Wide Quality Range: A Case Study}},
  author={Bhat, Madhukar and others},
  booktitle={{2021 IEEE Intl. Conf. on Image Processing}},
  year={2021},
  organization={{IEEE}}
}

@inproceedings{azimi2024decoding,
  title={{Decoding Complexity-Aware Bitrate-Ladder Estimation for Adaptive VVC Streaming}},
  author={Azimi, Zoha and others},
  booktitle={{32nd European Signal Processing Conference}},
  year={2024}
}

@inproceedings{mueller2022context,
  title={{Context-Aware Video Encoding as a Network-Based Media Processing Workflow}},
  author={Mueller, Christoph and others},
  booktitle={{Proc. of the 13th ACM Multimedia Systems Conference}}, 
  year={2022}
}

@INPROCEEDINGS{HLS_ladder_ref,
    author={{Apple Inc.}},
    title = {{HLS Authoring Specification for Apple Devices}},
    url = {https://developer.apple.com/documentation/http_live_streaming/hls_authoring_specification_for_apple_devices}
}

@misc{HLS,
  number=8216,
  howpublished={RFC 8216},
  publisher={RFC Editor},
  url={https://rfc-editor.org/rfc/rfc8216.txt},
  author={Roger Pantos and William May},
  title={{HTTP Live Streaming}},
  year=2017,
note = {{Retrieved: 2026-02-06}}
}

@INPROCEEDINGS{netflix_paper,
  author={De Cock, Jan and others},
  booktitle={{IEEE Intl. Conf. on Image Processing }}, 
  title={{Complexity-based Consistent-Quality Encoding in the Cloud}}, 
  year={2016}
  }

@inproceedings{VVdeC_ref,
  title={{Towards a Live Software Decoder Implementation for the Upcoming Versatile Video Coding Codec}},
  author={Wieckowski, Adam and others},
  booktitle={{IEEE Intl. Conf. on Image Processing }},
  year={2020},
  organization={{IEEE}}
}

@article{ghasempour2025real,
  title={{Real-Time Quality-and Energy-Aware Bitrate Ladder Construction for Live Video Streaming}},
  author={Ghasempour, Mohammad and others},
  journal={{IEEE Journal on Emerging and Selected Topics in Circuits and Systems}},
  year={2025},
  publisher={{IEEE}}
}

@inproceedings{zhao2024efficient,
  title={{Efficient Bitrate Ladder Construction for Per-Shot Adaptive Encoding}},
  author={Zhao, Yan and others},
  booktitle={{IEEE Intl. Conf. on Visual Communications and Image Processing}},
  year={2024},
  organization={{IEEE}}
}

@INPROCEEDINGS{rajendran2024,
  author={Rajendran, Prajit T and others},
  booktitle={2024 IEEE Intl. Conf. on Visual Communications and Image Processing}, 
  title={{Energy-Quality-aware Variable Framerate Pareto-Front for Adaptive Video Streaming}}, 
  year={2024}}

@inproceedings{premkumar2024quality,
  title={{Quality-Aware Dynamic Resolution Adaptation Framework for Adaptive Video Streaming}},
  author={Premkumar, Amritha and others},
  booktitle={{Proc. of the 15th ACM Multimedia Systems Conf.}},
  year={2024}
}

@misc{AppLogic25,
  publisher={{AppLogic Networks}},
  url={https://www.applogicnetworks.com/blog/the-2025-global-internet-phenomena-report},
  author={Kurt Rosenthal},
  title={{The 2025 Global Internet Phenomena Report}},
  year={2025},
note = {{Retrieved: 2026-02-06}}
}

@article{telili2023bitrate,
  title={{Bitrate Ladder Prediction Methods for Adaptive Video Streaming: A Review and Benchmark}},
  author={Telili, Ahmed and others},
  journal={arXiv preprint arXiv:2310.15163},
  year={2023}
}

@article{bentaleb2022bob,
  title={{BoB: Bandwidth Prediction for Real-Time Communications using Heuristic and Reinforcement Learning}},
  author={Bentaleb, Abdelhak and others},
  journal={{IEEE Trans. on Multimedia}},
  year={2022},
  publisher={{IEEE}}
}

@inproceedings{katsenou2024decoding,
  title={{Decoding Complexity-Rate-Quality Pareto-Front for Adaptive VVC Streaming}},
  author={Katsenou, Angeliki and others},
  booktitle={{IEEE Intl. Conf. on Visual Communications and Image Processing}},
  year={2024},
  organization={{IEEE}}
}

@inproceedings{menon2024convex,
  title={{Convex-hull Estimation using XPSNR for Versatile Video Coding}},
  author={Menon, Vignesh V and others},
  booktitle={{IEEE Intl. Conf. on Image Processing }},
  year={2024},
  organization={{IEEE}}
}

@inproceedings{kranzler2020decoding,
  title={{Decoding Energy Modeling for Versatile Video Coding}},
  author={Kr{\"a}nzler, Matthias and others},
  booktitle={{IEEE Intl. Conf. on Image Processing }},
  year={2020},
  organization={{{IEEE}}}
}

@inproceedings{boujida2023decoding,
  title={{Decoding Time Prediction for Versatile Video Coding}},
  author={Boujida, Hafssa and others},
  booktitle={{IEEE 25th Intl. Workshop on Multimedia Signal Processing}},
  year={2023},
  organization={IEEE}
}

@inproceedings{cabarat2024pictures,
  title={{Pictures Decoding Time Estimation for Low-Power VVC Software Decoding}},
  author={Cabarat, Pierre-Loup and others},
  booktitle={{32nd European Signal Processing Conference}},
  year={2024},
  organization={{IEEE}}
}

@article{barman2024bj,
  title={{Bj $\{$$\backslash$o$\}$ ntegaard Delta (BD): A Tutorial Overview of the Metric, Evolution, Challenges, and Recommendations}},
  author={Barman, Nabajeet and others},
  journal={arXiv preprint arXiv:2401.04039},
  year={2024}
}

@inproceedings{solomatine2004adaboost,
  title={{AdaBoost. RT: A Boosting Algorithm for Regression Problems}},
  author={Solomatine, Dimitri P and Shrestha, Durga L},
  booktitle={{IEEE Intl. Joint Conf. on Neural Networks}},
  year={2004},
  organization={{IEEE}}
}

@article{ke2017lightgbm,
  title={{LightGBM: A Highly Efficient Gradient Boosting Decision Tree}},
  author={Ke, Guolin and Meng, Qi and Finley, Thomas and Wang, Taifeng and Chen, Wei and Ma, Weidong and Ye, Qiwei and Liu, Tie-Yan},
  journal={{Advances in Neural Information Processing Systems}},
  year={2017}
}

@article{geurts2006extremely,
  title={{Extremely Randomized Trees}},
  author={Geurts, Pierre and Ernst, Damien and Wehenkel, Louis},
  journal={{Machine Learning}},
  year={2006},
  publisher={{Springer}}
}

@article{popescu2009multilayer,
  title={{Multilayer Perceptron and Neural Networks}},
  author={Popescu, Marius-Constantin and others},
  journal={{WSEAS Trans. on Circuits and Systems}}, 
  year={2009}
}

@article{scikit-learn,
  title={{Scikit-learn: Machine Learning in {P}ython}},
  author={Pedregosa, F. and others},
  journal={{Journal of Machine Learning Research}},
  year={2011}
}

@inproceedings{optuna_2019,
    title={{Optuna: A Next-generation Hyperparameter Optimization Framework}},
    author={Akiba, Takuya and others},
    booktitle={Proc. of the 25th {ACM} {SIGKDD} Intl. Conf. on Knowledge Discovery and Data Mining},
    year={2019}
}

@article{lundberg2017unified,
  title={{A Unified Approach to Interpreting Model Predictions}},
  author={Lundberg, Scott M and Lee, Su-In},
  journal={{Advances in Neural Information Processing Systems}},
  year={2017}
}

@article{bentaleb2025toward,
  title={{Toward One-Second Latency: Evolution of Live Media Streaming}},
  author={Bentaleb, Abdelhak and others},
  journal={IEEE Communications Surveys \& Tutorials},
  year={2025},
  publisher={IEEE}
}

@article{li2024perceptual,
  title={{Perceptual Quality Assessment of Face Video Compression: A Benchmark and an Effective Method}},
  author={Li, Yixuan and others},
  journal={IEEE Trans. on Multimedia},
  year={2024},
  publisher={IEEE}
}

@inproceedings{nair2010rectified,
  title={{Rectified Linear Units Improve Restricted Boltzmann Machines}},
  author={Nair, Vinod and Hinton, Geoffrey E},
  booktitle={{Proc. of the 27th Intl. Conf. on Machine Learning}},
  year={2010}
}

@inproceedings{he2015delving,
  title={{Delving Deep Into Rectifiers: Surpassing Human-Level Performance on Imagenet Classification}},
  author={He, Kaiming and others},
  booktitle={{Proc. of the IEEE Intl. Conf. on Computer Vision}},
  year={2015}
}

@article{zhang2022bilateral,
  title={{Bilateral Sensitivity Analysis: A Better Understanding of a Neural Network}},
  author={Zhang, Huaqing and others},
  journal={{Intl. Journal of Machine Learning and Cybernetics}},
  year={2022},
  publisher={{Springer}}
}

@article{wisniewski2025benchmarking,
  title={{Benchmarking Deep Reinforcement Learning for Navigation in Denied Sensor Environments}},
  author={Wisniewski, Mariusz and others},
  journal={{Journal of Intelligent \& Robotic Systems}},
  year={2025},
  publisher={{Springer}}
}

@article{amirpour2025vqm4has,
  title={{VQM4HAS: A Real-time Quality Metric For HEVC Videos in HTTP Adaptive Streaming}},
  author={Amirpour, Hadi and Zhu, Jingwen and Zhou, Wei and Le Callet, Patrick and Timmerer, Christian},
  journal={{IEEE Trans. on Multimedia}},
  year={2025},
  publisher={{IEEE}}
}

@article{lebreton2020predicting,
  title={{Predicting User Quitting Ratio in Adaptive Bitrate Video Streaming}},
  author={Lebreton, Pierre and Yamagishi, Kazuhisa},
  journal={{IEEE Trans. on Multimedia}},
  year={2020},
  publisher={{IEEE}}
}

@article{hands2004basic,
  title={{A Basic Multimedia Quality Model}},
  author={Hands, David S},
  journal={{IEEE Trans. on Multimedia}},
  year={2004},
  publisher={{IEEE}}
}
\end{document}